\documentclass[useAMS,usenatbib]{mn2e}
\usepackage{amssymb,epsfig,natbib}

\title[A Study of the Type II-P Supernova 2003gd in M74]
  {A Study of the Type II-P Supernova 2003gd in M74}
\author[M. A. Hendry {\rm et al.}]
{M. A. Hendry$^1$, S. J. Smartt$^{2}$, J. R. Maund$^1$, A. Pastorello$^3$, L. Zampieri$^4$, S. Benetti$^4$,  
\newauthor M. Turatto$^4$, E. Cappellaro$^5$, W. P. S. Meikle$^6$, R. Kotak$^6$, M. J. Irwin$^1$, P. G. Jonker$^7$,
\newauthor L. Vermaas$^8$, R. F. Peletier$^8$, H. van Woerden$^8$, K. M. Exter$^9$, D. L. Pollacco$^{10}$,
\newauthor S. Leon$^{11}$, S. Verley$^{11}$, C. R. Benn$^{12}$ and G. Pignata$^{4,13}$\\
\\
$^1$Institute of Astronomy, University of Cambridge, Madingley Road,
        Cambridge CB3 0HA, England, UK\\
$^2$Department of Physics and Astronomy, Queen's University Belfast, Belfast BT7 1NN, Northern Ireland, UK\\
$^3$INAF - Osservatorio Astrofisico di Arcetri, Largo E. Fermi 5, I-50125 Firenze, Italia\\
$^4$INAF - Osservatorio Astronomico di Padova, Vicolo dell' Osservatorio 5, I-35122 Padova, Italia\\
$^5$INAF - Osservatorio Astronomico di Capodimonte, Via Moiariello 16, I-80131 Napoli, Italia\\
$^6$Astrophysics Group, Blackett Laboratory, Imperial College London, Prince Consort Road, London SW7 2BZ, England, UK\\
$^7$Harvard-Smithsonian Center for Astrophysics, Cambridge, MA 02138, USA\\
$^8$Kapteyn Instituut, Postbus 800, 9700 AV Groningen, The Netherlands\\
$^9$Instituto de Astrof\'{\i}sica de Canarias
C/Via Lactea s/n, E38200 - La Laguna (Tenerife), Espa\~{n}a\\
$^{10}$School of Pure and Applied Physics, Queen's University Belfast, Belfast BT7 9NN, Northern Ireland, UK\\
$^{11}$Instituto de Astrof\'{\i}sica de Andaluc\'{\i}a (CSIC), Ap. 3004, Granada 18080, Espa\~{n}a\\
$^{12}$Isaac Newton Group, Apartado 321, E-38700 Santa Cruz de La Palma, Espa\~{n}a\\
$^{13}$European Southern Observatory (ESO), Karl-Schwarzschild-Str. 2, D-85748 Garching bei M\"{u}nchen, Germany}
\date{Released 2004 Xxxxx XX}

\pagerange{\pageref{firstpage}--\pageref{lastpage}} \pubyear{2004}

\def\LaTeX{L\kern-.36em\raise.3ex\hbox{a}\kern-.15em
    T\kern-.1667em\lower.7ex\hbox{E}\kern-.125emX}
\def \aj {AJ}
\def \mnras {MNRAS}
\def \apj {ApJ}
\def \apjl {ApJL}
\def \aap {A\&A}

\def \iauc {IAUC}
\def \pasp {PASP}

\def \apjs {ApJS}
\def \aaps {A\&AS}

\def \msun {M$_{\odot}$}
\def \hub {km s$^{-1}$ Mpc$^{-1}$}

\begin{document}

\label{firstpage}

\maketitle

\begin{abstract}
We present photometric and spectroscopic data of the type II-P supernova 2003gd, which was discovered in M74 close to the end of its plateau phase. SN~2003gd is the first type II supernova to have a directly confirmed red supergiant progenitor. We compare SN~2003gd with SN~1999em, a similar type II-P supernova, and estimate an explosion date of 18th March 2003. We determine a reddening towards the supernova of $E(B-V) = 0.14\pm0.06$, using three different methods. We also calculate three new distances to M74 of $9.6\pm2.8$ Mpc, $7.7\pm1.7$ Mpc and $9.6\pm2.2$ Mpc. The former was estimated using the Standardised Candle Method (SCM), for type II supernovae, and the latter two using the Brightest Supergiants Method (BSM). When combined with existing kinematic and BSM distance estimates, we derive a mean value of $9.3\pm 1.8$~Mpc. SN~2003gd was found to have a lower tail luminosity compared to other ``normal'' type II-P SNe bringing into question the nature of this supernova. We present a discussion concluding that this is a ``normal'' type II-P supernova which is consistent with the observed progenitor mass of $8^{+4}_{-2}$ \msun.
\end{abstract}


\section{Introduction}

Supernova (SN) 2003gd was discovered by Rev. Bob Evans on June 12.82 2003 UT (JD 2452803.32) with an apparent magnitude of $\sim$13.2 mag, situated on the southern spiral arm of M74. The SN was soon confirmed by R. H. McNaught on June 13.84 UT who gave a precise position of R. A. = $1^{h}36^{m}42^{s}.65$, Dec = +15$^{\circ}44\arcmin20\arcsec.9$ \citep{2003IAUC.8150....2E}. \citet{2003IAUC.8150....3G} obtained a near-infrared spectrum of the SN on June 13.46 UT showing broad Pa$\alpha$ and Pa$\gamma$ which are indicative of a supernova of type II. \citet{2003IAUC.8152....1K} confirmed 2003gd as a type II with an optical spectrum and estimated that it was roughly 2 months post-explosion. Owing to the relative proximity of M74 (9.3 Mpc; see \S\ref{sec:D}) and its favourable inclination, a monitoring programme was initiated. M74 was also the host galaxy of the recent type Ic SN~2002ap \citep{2002ApJ...572L..61M,2002ApJ...572L.147S,2002MNRAS.332L..73G}. \citet{2003PASP..115.1289V} presented early $BVRI$ light curves and colour curves of SN~2003gd and compared them with those of SN~1999em. They photometrically classified 2003gd as a type II-P SN and also gave estimates of the explosion date, 17th March 2003, and of the line of sight reddening, $E(B-V) = 0.13\pm 0.03$. Using ground based astrometric calibrations \citet{2003IAUC.8152....4S} and \citet{2003PASP..115.1289V} estimated the position of SN~2003gd on pre-discovery Hubble Space Telescope ({\it HST}) images of M74, suggesting possible candidates for the progenitor but both studies required follow-up {\it HST} images to determine the SN position with sufficient precision to unambiguously identify any precursor star.

After observing SN~2003gd with the Advanced Camera for Surveys (ACS) as part of our ongoing {\it HST} programme on supernova progenitors, we presented the discovery of a star which exploded as SN~2003gd \citep{2004Sci...303..499S}. The progenitor star was also identified on pre-explosion Gemini images and the combined $VRI$ magnitudes were used to estimate its intrinsic colour and luminosity. These were found to be consistent with those of a red supergiant, and comparison with stellar evolutionary tracks suggested that it had an initial main sequence mass of $8^{+4}_{-2}$ \msun. This is only the third time that a progenitor of a genuinely confirmed SN has been identified and is the first apparently normal red supergiant to be associated with a SN of type II-P. The determination of intrinsic colour and luminosity require the reddening and distance to be known. We summarised the results of our measurements for both of these quantities in \citet{2004Sci...303..499S} and in this paper we present the full photometric and spectroscopic data set. Given the discovery of a red supergiant progenitor star, which have long been assumed to be the progenitors of SNe~II-P, it is essential that the SN itself is studied and quantified. 

In this paper we present extensive photometric and spectroscopic data of SN~2003gd in \S\ref{sec:obs} followed by an analysis of the photometry in \S\ref{sec:epoch}, where an explosion date is estimated. We estimate the reddening towards the supernova, using three different methods, in \S\ref{sec:red} and analyse the velocity evolution in \S\ref{sec:expvel}. The distance to M74 is not well known and as yet no Cepheid distance exists. We attempt to improve the situation in \S\ref{sec:Dw} by estimating the distance using two different methods and compiling all distances within the literature. Using the distance found we then calculate the amount of $^{56}$Ni synthesised in the explosion in \S\ref{sec:Ni}. In \S\ref{sec:zamp} the observed properties of SN~2003gd are compared to the semi-analytical model of \citet{2003MNRAS.338..711Z} and in \S\ref{sec:diss} we discuss the nature of the supernova and the implication for the progenitor. We then conclude in \S\ref{sec:con}. Throughout this work we have assumed the galactic reddening laws of \citet*{1989ApJ...345..245C} with $R_V = 3.1$.

\section{Observations}\label{sec:obs}

\subsection{Photometry of SN~2003gd}
\label{sec:opphot}

$BVRI$ photometry was obtained shortly after discovery and covers a range of $\sim$92-169 days after explosion with a few late-time epochs around 490 days. These data were collected from several telescopes: the 1.0 m Jacobus Kapteyn Telescope (JKT), the 2.5 m Isaac Newton Telescope (INT) and the 3.5 m Telescopio Nazionale Galileo (TNG), all on La Palma, and the 1.82 m Copernico Telescope (CTA) at Asiago, Italy. The latest data were taken with Unit Telescope 1 (UT1), Antu, of the ESO Very Large Telescope (VLT), Paranal, Chile. A summary of these observations can be found in Table \ref{tab:opphot} as well as the results from the optical photometry.

The JKT observations where taken with its $2048\times2048$ SITe1 JAG-CCD camera and were reduced using standard techniques within {\sc iraf}. The frames were debiased and flat-fielded using twilight sky flats generally taken on the night of the observations. In some cases flats for our required filters were not available from that night and we employed the best set of flats closest in time to the supernova observations. An extensive photometric sequence of stars around M74 has been calibrated by \citet{2002GCN..1242....1H} originally intended for monitoring of SN~2002ap.  On the JKT images there were normally nine stars suitable for calibration of the supernova magnitudes (Henden numbers 100, 109, 149, 153, 156, 158, 160, 161) and these were used to determine the zero-points and transformation coefficients on each night. In some cases the telescope pointing was slightly different or the brightest stars in the sequence were saturated due to increased exposure time for the fading supernova and hence not all the standards could be used. A minimum of three standard stars were always available on each night. The $BVRI$ magnitudes were obtained using aperture photometry and the errors are the standard deviations of the individual values of the supernova magnitudes determined from each calibration star.

The INT observations were taken with the Wide Field Camera (WFC) at the prime focus. The WFC consists of 4 thinned EEV 2k$\times$4k CCDs and SN~2003gd and its calibration stars were placed on CCD4 of the mosaic. The data were reduced using the standard pipeline processing applied to all WFC images in the Cambridge archive \citep{2001NewAR..45..105I}. Images were taken with the Harris $V$ filter and the SDSS-like filters $r'$ and $i'$. We attempted to transform the $r'$ and $i'$ aperture magnitudes to the $RI$ system of Henden using 15 stars from the Henden catalogue. We derived good fits between the instrumental magnitudes and the intrinsic $(R-I)$ and $(V-R)$ colours for the stars and tested these on four other stars in common to the Henden and WFC catalogues. The transformations produced $VRI$ magnitudes which were accurate to a mean of 0.03. However when we transformed the supernova CCD $r'i'$ magnitudes to the $RI$ system we got very unsatisfactory results. On 2003 August 25, we derived $R=16.20$ and $I=16.34$, which are very discrepant from our AFOSC magnitudes taken the previous night. As the transformations for field stars proved to be reliable the only explanation for this serious discrepancy is the strong emission line SN spectrum producing systematic differences in the colour terms compared to the stellar SEDs. Hence we quote the $V$ band magnitude only for these two nights as the $r'i'$ transformations cannot be relied upon to produce magnitudes consistent with our previous calibrations. The errors in the magnitudes were calculated as with the JKT data.

\begin{figure}
\begin{center}
\epsfig{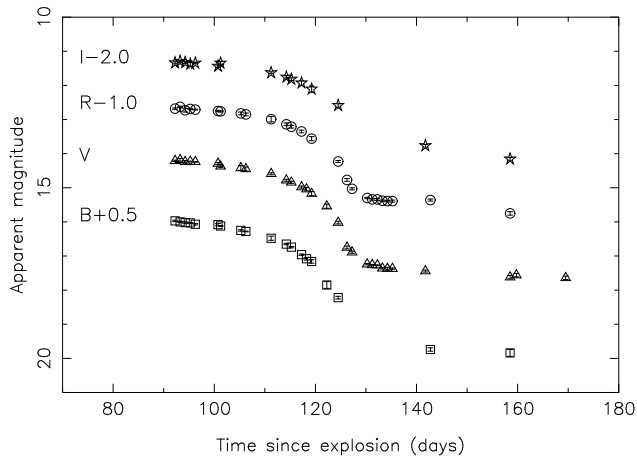}
\caption{$BVRI$ light curves of SN~2003gd which have been arbitrarily shifted in magnitude for clarity.}
\label{fig:BVRI}
\end{center}
\end{figure}

The CTA observations were taken with the Asiago Faint Object Spectrograph and Camera (AFOSC), a focal reducer instrument, which is equipped with a Tektronix 1024$\times$1024 CCD with a pixel sampling of 0.47 arcsec pixel$^{-1}$. The TNG observations were instead secured with the Device Optimized for LOw RESolution (DOLORES), which is equipped with a Loral 2k$\times$2k CCD with a pixel scale of 0.275 arcsec pixel$^{-1}$. The data were reduced using standard techniques within {\sc iraf}. The SN magnitudes were measured using the {\sc iraf} point-spread-function (PSF) fitting task {\sc daophot}. This procedure allows the simultaneous fitting and subtraction of the galaxy background. For cases in which the seeing is fair, where the SN is relatively bright and its PSF well-sampled, it has been found that this method produces results in excellent agreement with the template subtraction method \citep[see][]{2003MNRAS.340..191R}. The errors in the derived magnitudes were calculated by combining in quadrature the standard deviation of the magnitudes found from each of the calibration stars, the PSF fitting error and the standard deviation of the magnitudes of artificial stars placed around the SN position having the SN magnitude. The photometry results, from all telescopes, are tabulated in Table \ref{tab:opphot}. The earlier data are plotted in the $BVRI$ light curves in Figure~\ref{fig:BVRI} whereas the late-time data are shown in Figure~\ref{fig:LClt}.

\begin{table*}
\caption[]{Journal and results of optical photometry of SN~2003gd}
\begin{minipage}{\linewidth}
\begin{center}
\begin{tabular}{lrrrrrrll} \hline
Date & JD &  Phase & $B$ & $V$ & $R$ & $I$ & Telescope +& Observer \\
 & (2450000+) & (days) & & & & &Instrument &\\
\hline
2003 Jun 19 & 2809.20 & 92  & 15.47(0.02) & 14.21(0.02) & 13.68(0.02) &
13.34(0.03) & JKT+JAG & Jonker\\ 
2003 Jun 20 & 2810.21 & 93  & 15.50(0.02) & 14.18(0.04) & 13.63(0.02) &
13.30(0.04) & JKT+JAG & Jonker\\
2003 Jun 21 & 2811.21 & 94  & 15.52(0.02) & 14.24(0.04) & 13.73(0.05) &
13.33(0.03) & JKT+JAG & Jonker\\
2003 Jun 22 & 2812.21 & 95  & 15.53(0.01) & 14.23(0.01) & 13.69(0.02) &
13.38(0.05) & JKT+JAG & Jonker\\
2003 Jun 23 & 2813.20 & 96  & 15.57(0.02) & 14.24(0.01) & 13.71(0.01) &
13.35(0.02) & JKT+JAG & Jonker\\
2003 Jun 27$^{a}$ & 2817.66 & 100  & 15.58(0.03) & 14.29(0.03) &
13.75(0.02) & 13.44(0.04) & TNG+DOLORES & Zampieri\\
2003 Jun 28 & 2818.19 & 101  & 15.62(0.01) & 14.37(0.04) & 13.76(0.02) &
13.35(0.05) & JKT+JAG & Vermaas\\
2003 Jul 02 & 2822.20 & 105  & 15.75(0.02) & 14.42(0.07) & 13.82(0.04) & - &
JKT+JAG & Vermaas\\
2003 Jul 03 & 2823.19 & 106  & 15.78(0.01) & 14.45(0.06) & 13.85(0.04) & - &
JKT+JAG & Vermaas\\
2003 Jul 08 & 2828.18 & 111  & 15.98(0.06) & 14.59(0.03) & 13.99(0.08) &
13.63(0.03) & JKT+JAG & Exter\\
2003 Jul 11 & 2831.21 & 114  & 16.15(0.02) & 14.78(0.03) & 14.14(0.04) &
13.76(0.05) & JKT+JAG & Exter\\
2003 Jul 12 & 2832.22 & 115  & 16.24(0.02) & 14.84(0.03) & 14.21(0.04) &
13.82(0.04) & JKT+JAG & Exter\\
2003 Jul 14 & 2834.22 & 117  & 16.46(0.02) & 14.98(0.03) & 14.35(0.04) &
13.93(0.06) & JKT+JAG & Exter\\
2003 Jul 15 & 2835.15 & 118  & 16.58(0.04) & 15.04(0.05) & - & - & JKT+JAG &
Exter\\
2003 Jul 16 & 2836.23 & 119  & 16.66(0.05) & 15.17(0.06) & 14.56(0.06) &
14.11(0.09) & JKT+JAG & Exter\\
2003 Jul 19 & 2839.23 & 122  & 17.35(0.10) & 15.54(0.06) & - & - &
JKT+JAG & Exter\\
2003 Jul 21 & 2841.56 & 124  & 17.72(0.04) & 16.02(0.04) &
15.23(0.03) & 14.59(0.05) & CTA+AFOSC & Pastorello\\
2003 Jul 23 & 2843.18 & 126  & - & 16.75(0.04) & 15.77(0.04) & - & JKT+JAG &
Verley/Leon\\
2003 Jul 24 & 2844.19 & 127  & - & 16.89(0.03) & 16.03(0.03) & - &
JKT+JAG & Verley/Leon\\ 
2003 Jul 26 & 2847.19 & 128  & - & 17.24(0.02) & 16.30(0.02) & - &
JKT+JAG & Verley/Leon\\ 
2003 Jul 27 & 2848.18 & 129  & - & 17.26(0.01) & 16.34(0.02) & - &
JKT+JAG & Verley/Leon\\ 
2003 Jul 28 & 2849.18 & 130  & - & 17.27(0.01) & 16.34(0.03) & - &
JKT+JAG & Verley/Leon\\ 
2003 Jul 29 & 2850.18 & 131  & - & 17.36(0.02) & 16.38(0.02) & - &
JKT+JAG & Verley/Leon\\ 
2003 Jul 30 & 2851.18 & 132  & - & 17.37(0.04) & 16.39(0.02) & - &
JKT+JAG & Verley/Leon\\ 
2003 Jul 31 & 2852.18 & 133  & - & 17.37(0.02) & 16.39(0.03) & - &
JKT+JAG & Verley/Leon\\
2003 Aug 07 & 2858.73 & 141  & - & 17.44(0.03) & - & 15.77(0.04) &
TNG+DOLORES & Zampieri\\
2003 Aug 08$^{b}$ & 2859.72 & 142  & 19.24(0.05) & - & 16.36(0.03) & - &
TNG+DOLORES & Zampieri\\
2003 Aug 24 & 2875.51 & 158  & 19.34(0.10) & 17.62(0.05) &
16.75(0.05) & 16.16(0.05) & CTA+AFOSC & Benetti\\
2003 Aug 25 & 2876.71 & 159  & - & 17.56(0.05) & - & - & INT+WFC & Irwin\\
2003 Sep 04 & 2886.61 & 169  & - & 17.64(0.05) & - & - & INT+WFC & Irwin\\
2004 Jul 15 & 3201.69 & 485  & - & - & 20.55(0.09) & - & TNG+DOLORES & Zampieri\\
2004 Jul 23 & 3209.65 & 493 & - & 21.33(0.26) & - & 20.30(0.24) & TNG+DOLORES & Zampieri\\ 
2004 Jul 23 & 3209.91 & 493 & 21.76(0.06) & 21.32(0.07) & 20.59(0.09) & 20.31(0.04) & VLT-UT1+FORS2 & Patat/Pignata \\
\hline
\end{tabular}\\
\end{center}
{\scriptsize NOTE:- Figures in brackets give the errors associated with the magnitudes. A full account of these errors can be found in \S\ref{sec:opphot}.\\
$^a$ $U = 17.19$, $^b$ $U = 21.39$\\
JKT = 1.0 m Jacobus Kapteyn Telescope, La Palma\\
TNG = 3.5 m Telescopio Nazionale Galileo, La Palma\\
CTA = 1.82 m Copernico Telescope, Asiago, Italy\\
INT = 2.5 m Isaac Newton Telescope, La Palma\\
VLT-UT1 = 8.2 m ESO Very Large Telescope, Antu (UT1), Paranal, Chile}
\end{minipage}
\label{tab:opphot}
\end{table*}

\subsection{Spectroscopy of SN~2003gd}\label{sec:spec}

\begin{figure*}
\begin{center}
\epsfig{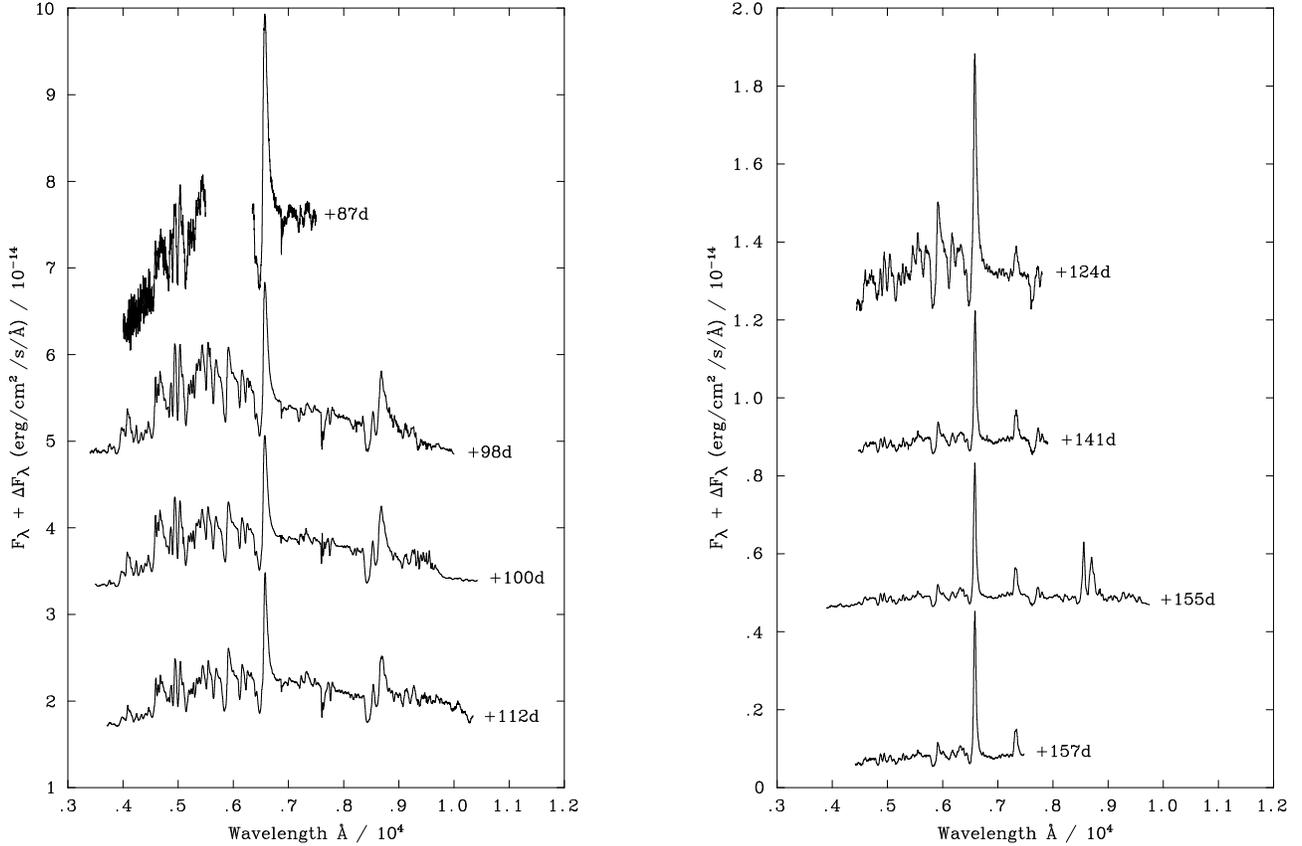}
\caption{Spectral evolution of SN~2003gd from 87-157 days after explosion, where the spectra have been arbitrarily shifted in flux for clarity.} 
\label{fig:2003gd_specev}
\end{center}
\end{figure*}

\begin{figure*}
\centering
\epsfig{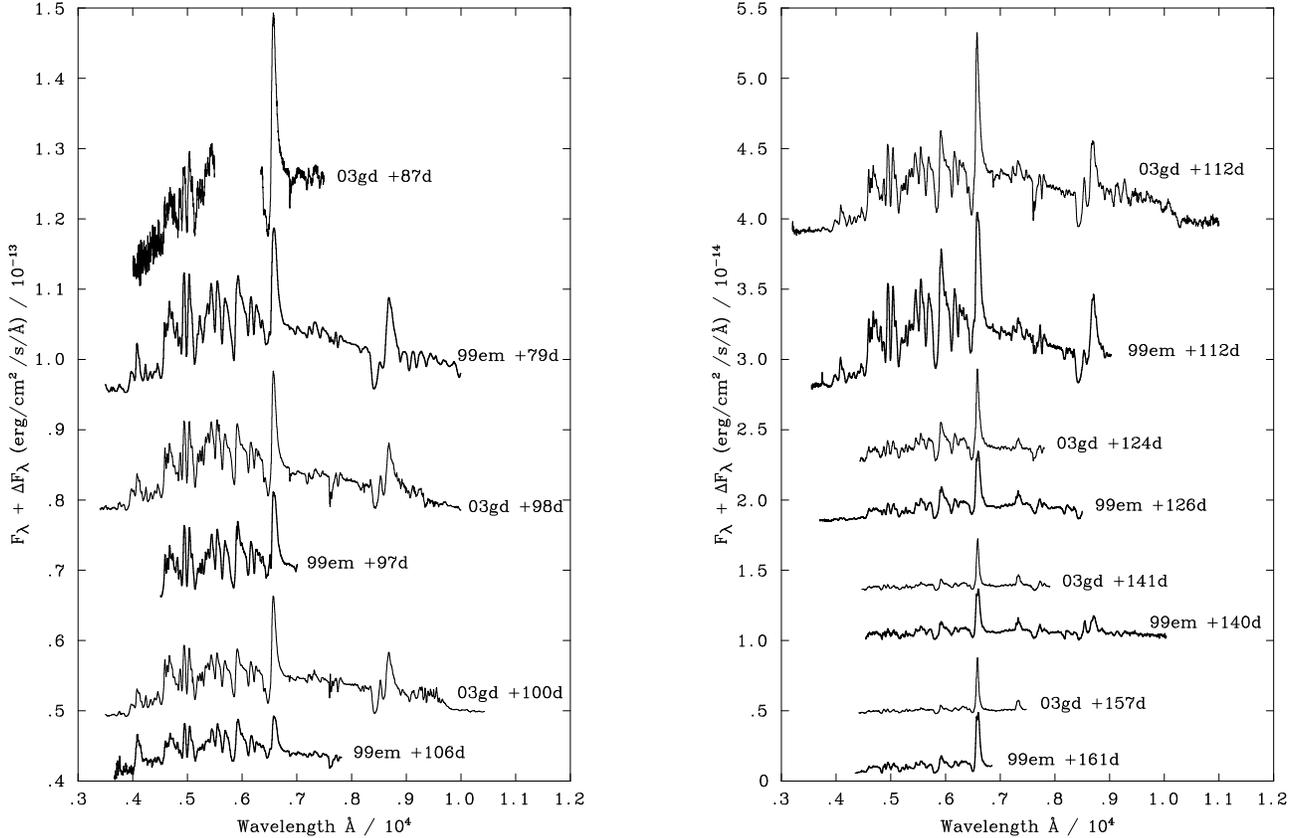}
\caption{Comparison of the spectra of SN~2003gd and SN~1999em at similar phases, where the spectra have been arbitrarily shifted in flux for clarity. The spectra of both supernovae show remarkable similarities at all epochs.} 
\label{fig:03gd99em}
\end{figure*}

Optical spectroscopic observations were taken at five telescopes and cover eight epochs, which we estimate are between 87 and 157 days after explosion, and one late-time epoch at 493 days. The log of the observations is shown in Table \ref{tab:spec}. The data were reduced using standard routines within {\sc iraf}. The frames were debiased, flat-fielded and extracted and then wavelength calibrated using Cu-Ar and Cu-Ne lamp spectra. The wavelength calibration was checked by determining the positions of the night sky lines and small adjustments were made. The spectra were then flux calibrated using spectrophotometric flux standards observed with the same instrumental setup. The slit-widths employed were generally between $1-2.1$ arcsec, hence the flux calibration is unlikely to give an accurate absolute scale.  We used the $BVRI$ photometry to adjust the absolute scale of the spectra at each epoch. Spectral $BVRI$ magnitudes were calculated from the spectra using the {\sc iraf} package {\sc synphot} within {\sc stsdas}. These magnitudes were compared to the observed photometric magnitudes and a scaling factor was calculated. The spectra were then read into the spectral analysis program {\sc dipso} \citep*{SUN50.24} for further analysis. The earlier spectral evolution of SN~2003gd is shown in Figure \ref{fig:2003gd_specev} and is compared to that of the well observed SN~1999em in Figure \ref{fig:03gd99em}. The late-time spectrum is shown in Figure \ref{fig:ltspec} of \S\ref{sec:ltspecphot}. These spectra will be available through the {\sc suspect}\footnote{http://bruford.nhn.ou.edu/$\sim$suspect/} website.

\begin{table*}
\caption[]{Journal of spectroscopic observations of SN~2003gd}
\begin{minipage}{\linewidth}
\begin{center}
\begin{tabular}{lrrrrlll} \hline
Date & JD & Phase & Range & Resolution & Instrument & Telescope & Observer \\
& (2450000+) & (days) & (\AA) &  (\AA) & & & \\
\hline
2003 Jun 14 & 2804.72 & 87 & 3953-7499 & 4 & ISIS & WHT & Benn\\
2003 Jun 25 & 2815.68 & 98 & 3402-9998 & 8 & IDS & INT & Lennon\\
2003 Jun 27 & 2817.67 & 100 & 3500-10430 & $\sim$14 & DOLORES & TNG & Zampieri\\
2003 Jul 09 & 2829.68 & 112 & 3200-11000 & 8 & IDS & INT & Prada\\
2003 Jul 21 & 2841.60 & 124 & 3494-7809 & $\sim$24 & AFOSC & CTA & Pastorello\\
2003 Aug 07 & 2858.71 & 141 & 3156-8066 & $\sim$14 & DOLORES & TNG & Zampieri\\
2003 Aug 22 & 2873.72 & 155 & 3147-10440 & $\sim$14 & DOLORES & TNG & Zampieri\\
2003 Aug 24 & 2875.55 & 157 & 3486-7475 & $\sim$24 & AFOSC & CTA & Benetti\\
2004 Jul 23 & 3209.90 & 493 & 4233-9632 & $\sim$13 & FORS2 & VLT-UT1 & Patat/Pignata\\
\hline
\end{tabular}\\
\end{center}
{\scriptsize WHT = 4.2 m William Herschel Telescope, La Palma\\
TNG = 3.5 m Telescopio Nazionale Galileo, La Palma\\
CTA = 1.82 m Copernico Telescope, Asiago, Italy\\
INT = 2.5 m Isaac Newton Telescope, La Palma\\
VLT-UT1 = 8.2 m ESO Very Large Telescope, Antu (UT1), Paranal, Chile}
\end{minipage}
\label{tab:spec}
\end{table*}


\subsection{{\it HST} ACS and WFPC2 Observations of M74}

\subsubsection{ACS HRC Imaging of SN~2003gd}\label{sec:acs}

$BV$ and $I$ imaging of SN~2003gd was acquired with the High Resolution Camera (HRC) of the Advanced Camera for Surveys (ACS) on {\it HST}.  SN~2003gd was observed at an epoch $\sim$137d post explosion as part of program GO9733.  These observations were acquired in order to image the surroundings of the supernova to determine a value of the reddening towards stars near the supernova and to determine the exact location of the supernova with respect to nearby stars.  These observations allowed the location of the supernova to be determined on pre-explosion observations and a direct identification of the progenitor star \citep{2004Sci...303..499S}. Details of the {\it HST} ACS observations are given in Table \ref{tab:tab1}.

\begin{table*}
\caption{Summary of {\it HST} observations of SN~2003gd and M74.}
\label{tab:tab1}
\begin{tabular}{lllrl}
\hline
Date & Filter &  Dataset & Exposure (s) & Instrument\\
\hline
2003 Aug 1  & F439W  & J8NV01011/21 & 2500 &  ACS/HRC\\
2003 Aug 2  & F555W  & J8NV01031/41 & 1100 &  ACS/HRC\\     
2003 Aug 2  & F814W  & J8NV01051    & 1350 &  ACS/HRC\\
\\                                                             
2001 Oct 3  & F450W  & U6EA0101/02  & 460  &  WFPC2  \\
2001 Oct 3  & F814W  & U6EA0103/04  & 460  &  WFPC2  \\ 
\hline
\end{tabular}
\end{table*}

The on-the-fly re-calibrated (OTFR) ACS images were obtained from the Space Telescope European Coordinating Facility archive.  Photometry was conducted on these frames using the {\sc iraf} package {\sc daophot} \citep{1987PASP...99..191S} and its incorporated PSF fitting algorithm {\sc allstar} \citep{1988AJ.....96..909S}. {\it HST} photometric system magnitudes, of stars and the supernova in these images, were converted to the Johnson-Cousins' system using the transformations of \citet{1995PASP..107.1065H}.

\subsubsection{WFPC2 Observations of M74}\label{sec:wfpc2}

M74 was imaged, in $B$ and $I$, with the Wide Field Planetary Camera 2 (WFPC2) instrument on {\it HST} as part of program GO9042 although these observations did not include the location of SN~2003gd.  Details of these observations are given in Table \ref{tab:tab1}.  The OTFR images were retrieved from the Space Telescope European Coordinating Facility archive, with cosmic ray split frames pre-combined.  Photometry was conducted on these frames using the HSTphot version 1.15b package \citep{2000PASP..112.1397D,2000PASP..112.1383D}. HSTphot includes corrections for charge transfer efficiency, PSF variation and aperture size, and converts instrumental magnitudes to standard $UBVRI$ magnitudes automatically. Stars, suitable for the brightest supergiants distance determination technique (see \S\ref{sec:BSGs}), were selected from the photometry output using the HSTphot object type classification scheme. Objects which were classified by HSTphot as extended, blended or containing bad pixels were discarded, as were probably blended stars or stars in crowded areas with PSF fit $\chi^{2}>2.5$.

\section{Analysis}

\subsection{Explosion epoch}\label{sec:epoch}

\begin{figure*}
\begin{center}
\epsfig{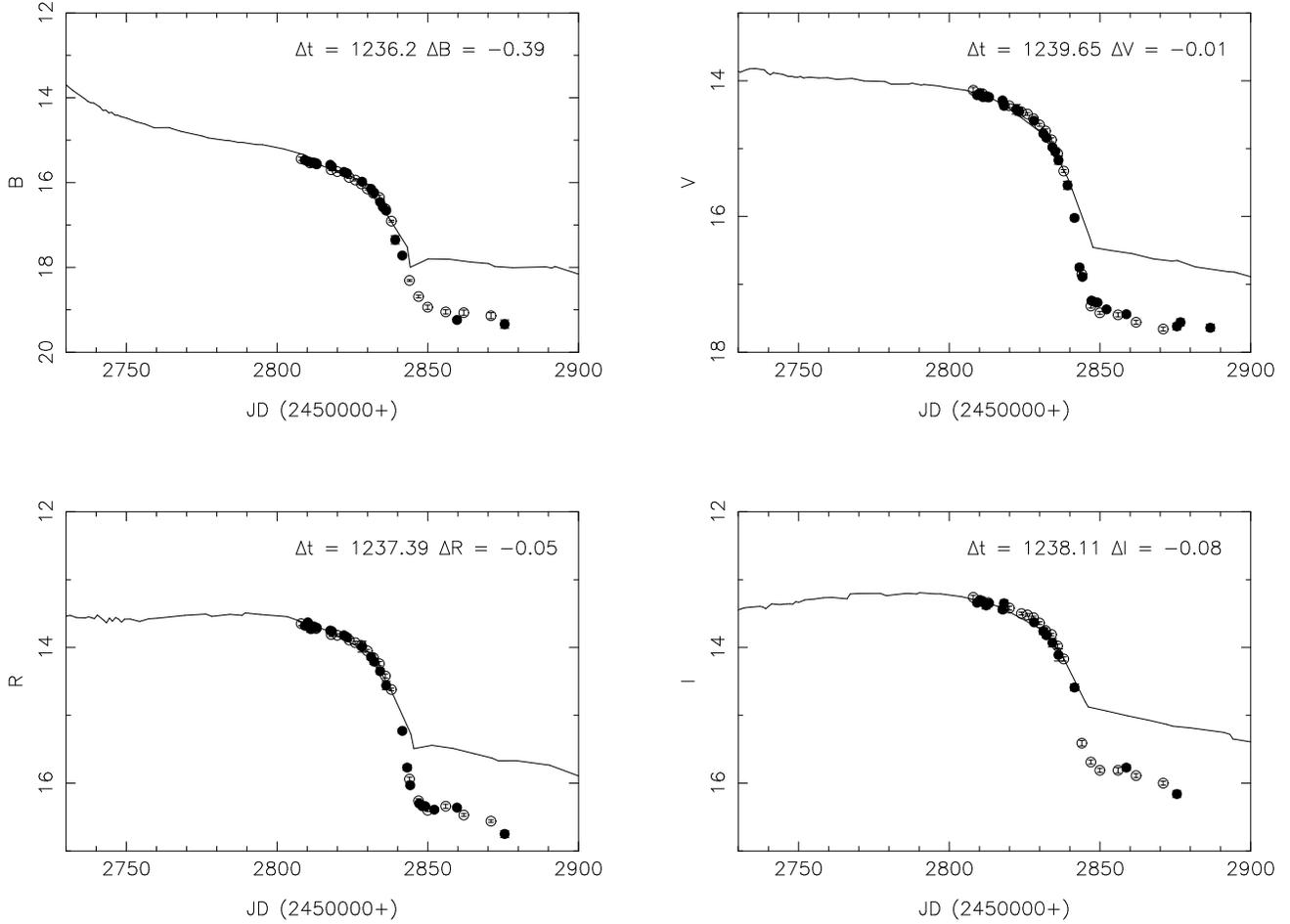}
\caption{$BVRI$ light curves of SN~2003gd over-plotted with the light curve for SN~1999em from \citet{2001ApJ...558..615H}. Filled circles are from data in Table \ref{tab:opphot} and open circles are from \citet{2003PASP..115.1289V} for comparison. The reduced-$\chi^2$ of the fit for $BVRI$ are as follows: 6.92, 1.37, 1.02 and 0.69.} \label{fig:03gd99emLC}
\end{center}
\end{figure*}

The first photometric data were obtained $\sim$6 days after discovery. As can be seen from the $BVRI$ light curves (Figure \ref{fig:BVRI}) SN~2003gd was discovered close to the end of its plateau phase when the host galaxy became visible after its conjunction with the sun. The plateau phase ends $\sim$30 days later when the nebular tail phase begins. The light curve of SN~2003gd was compared to that of SN~1999em, a similar type II-P supernova, using a $\chi^2$-fitting algorithm to adjust the time and apparent magnitude to find the best fit. The SN~2003gd data from after JD 2452845 were not included in the fit due to the difference in tail luminosities. The $BVRI$ best fits are shown in Figure \ref{fig:03gd99emLC} with the shift in time and apparent magnitude inset in each figure. The results from the $\chi^2$-fitting algorithm are shown in Table \ref{tab:expchi2} as well as the weighted average of the shift in days for the $VRI$ bands.
\begin{table*}
\begin{center}
\caption{Results from the $\chi^2$-fitting algorithm which adjusts the time, $\Delta t$, and apparent magnitude, $\Delta m$, of the ``model'' light curve (SN~1999em) to find the best fit to the data points of SN~2003gd, where $\nu$ = number of data points - number of degrees of freedom.}
\label{tab:expchi2}
\begin{tabular}{lrrrr}\hline
Filter & Reduced-$\chi^2$ & $\nu$ & $\Delta t$ (days) & $\Delta m$\\
\hline
$B$ & 6.92 & 14 & 1236.20 & -0.39\\
$V$ & 1.37 & 12 & 1239.65(0.53) & -0.01(0.02)\\
$R$ & 1.02 & 12 & 1237.39(0.65) & -0.05(0.02)\\
$I$ & 0.69 & 10 & 1238.11(1.14) & -0.08(0.03)\\
\hline
$VRI$ &    &      & 1238.67(0.39) & \\
\hline
\end{tabular}
\end{center}
\end{table*}
Although $B$ has a large reduced-$\chi^2$, the fit to the ``knee'' in the light curve looks as good as the others. The poor fit could be due to one of two reasons: i) under estimation of the errors in the photometry or ii) the errors being non-Gaussian due to the lower magnitude in the $B$-band. The $B$-band data was subsequently removed from the explosion date calculation, but had it been included there would have been little difference in the result. The errors in $\Delta t$, the shift in days, and $\Delta m$, the shift in magnitude, are the errors determined from the $\chi^2$-fit at the 1$\sigma$ level and are shown in brackets in Table \ref{tab:expchi2}. The explosion date was determined using the weighted average of $\Delta t$, in $VRI$, and the explosion epoch of SN~1999em \citep{2001ApJ...558..615H}, which was estimated as JD 2451478.8$\pm$0.5, within two days of \citet{2003MNRAS.338..939E}. We find the explosion date of SN~2003gd therefore to be JD 2452717$\pm$21, which corresponds to 18th March 2003. The error in the explosion epoch was estimated from the error in the weighted average of $\Delta t$, the error in the SN~1999em explosion date and the uncertainty in the duration of the plateau of SN~2003gd. This latter uncertainty was estimated from the observed parameters of a sample of 13 SNe~II-P from \citet[Table 3]{2003ApJ...582..905H} where the mean plateau duration was found to be 131$\pm$21 days. The most significant error is introduced by the assumption that the plateau of SN~2003gd was of a similar duration to SN~1999em. We have based this assumption on the spectral similarity of the two events (see Figure \ref{fig:03gd99em} and \S\ref{sec:SCM}) and their comparable masses \citep{2002ApJ...565.1089S,2003MNRAS.338..939E,2004Sci...303..499S}. The error of $\pm21$ days amply accounts for any reasonable intrinsic difference that may exist. Given this explosion epoch we estimate that the supernova was discovered $\sim$86 days after explosion. The light curves of both supernovae are comparable during the plateau phase, but SN~2003gd is substantially fainter than SN~1999em in the late-time nebular phase. This is most likely due to a lower $^{56}$Ni mass synthesised in the explosion, which was also suggested by \citet{2003PASP..115.1289V}. 

\subsection{Reddening Estimation towards SN~2003gd}\label{sec:red}

The reddening was estimated towards SN~2003gd using three different methods. It was first of all estimated using the colours of the supernova compared to SN~1999em; secondly using {\it HST} photometry of the surrounding stars and finally from the reddening towards nearby H{\sc ii} regions.

\subsubsection{Comparison of the colour evolution of SN~2003gd}
\label{sec:redSN}

This method assumes that SNe~II-P all reach the same intrinsic colour towards the end of the plateau phase. This is, in turn, based on the assumption that the opacity of SNe~II-P is dominated by electron scattering and, therefore, the supernovae should reach the temperature of hydrogen recombination at this point \citep{1996ApJ...466..911E,2004mmu..sympE...2H}. As shown in \S\ref{sec:spec} the two supernovae are strikingly similar at all epochs of our sample. It is therefore reasonable to assume that SN~2003gd would indeed reach the same colour as SN~1999em. \citet{2004mmu..sympE...2H} investigated this method using a sample of 24 SNe~II-P with $BVI$ photometry using SN~1999em as the comparison supernova. \citet{2004mmu..sympE...2H} found serious discrepancies between the results obtained from $(B-V)$ and $(V-I)$ colours and notes that this is an unsatisfactory technique and other methods should be explored in future.

\begin{figure}
\centering
\epsfig{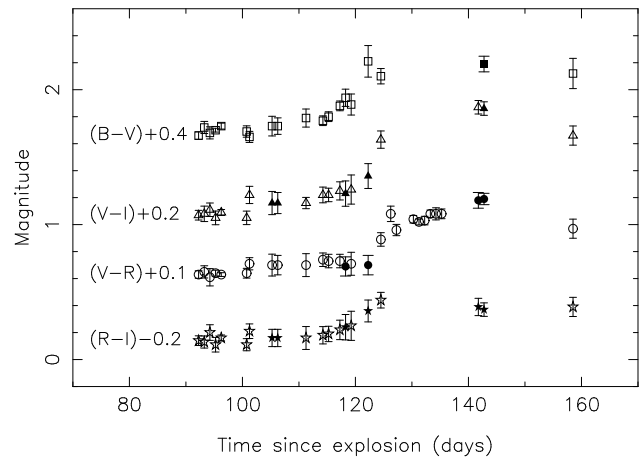}
\caption{Observed $(B-V)$, $(V-R)$, $(V-I)$ and $(R-I)$ colour curves of SN~2003gd plotted against the time since explosion. Open circles show direct observed colours whereas closed symbols are colours determined from the interpolated light curve.}
\label{fig:colour}
\end{figure}

\begin{figure}
\begin{center}
\epsfig{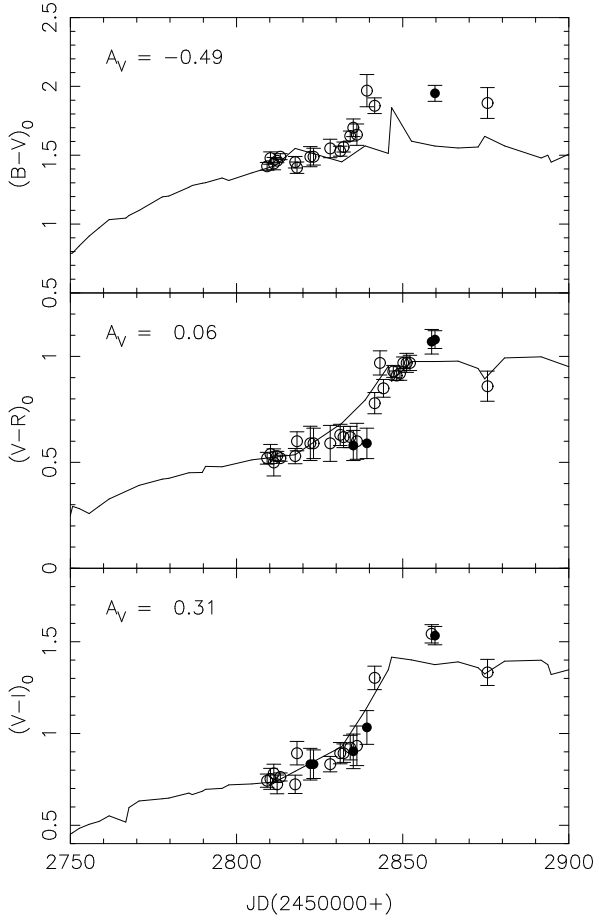}
\caption{$(B-V)$, $(V-R)$ and $(V-I)$ colour evolutions of SN~2003gd compared with the intrinsic colour evolutions of SN~1999em \citep{2001ApJ...558..615H}. The visual extinction required for each fit are given in the panels. The reduced-$\chi^2$ for $(B-V)$, $(V-R)$ and $(V-I)$ are 2.20, 0.27 and 1.01, respectively. Open circles show direct observed colours whereas closed symbols are colours determined from the interpolated light curve.}
\label{fig:colour03gd_99em}
\end{center}
\end{figure}

Bearing this in mind we carried out the analysis using $(B-V)$, $(V-R)$ and $(V-I)$. These colour curves for SN~2003gd can be seen in Figure \ref{fig:colour}.
SN~1999em was extensively modelled by \citet{2000ApJ...545..444B} who determined a hard upper limit of $E(B-V) \leq 0.15$ and estimated a likely reddening of $E(B-V) \simeq 0.05-0.10$. In this paper we adopt a reddening of $E(B-V) = 0.075\pm0.025$ for SN~1999em. The colour curves of SN~1999em were firstly dereddened using this value and were then shifted in time using the weighted average of $\Delta t$ discussed in \S\ref{sec:epoch}. A $\chi^2$-fitting algorithm was then used to compare the colours of SN~2003gd with those of SN~1999em for JD $<$ 2452830, the end of the plateau phase. The results of the $\chi^2$-fit are listed in Table \ref{tab:EBV} and shown in Figure \ref{fig:colour03gd_99em}. The errors in the colour excess, shown in brackets, were estimated from the $\chi^2$-fit at the 1$\sigma$ level. The systematic error, that the uncertainty in the reddening of SN~1999em introduces, was excluded from the fit to be added later.
\begin{table*}
\centering
\caption[]{Results of $\chi^2$-fitting algorithm to determine the reddening of SN~2003gd, where $\nu$ = number of data points - number of degrees of freedom.}
\begin{tabular}{lllrrr}\hline
Colour & Reduced-$\chi^2$ & $\nu$ & Colour Excess & $A_V$ & $E(B-V)$\\
\hline
$B-V$ & 2.20 & 9 & -0.16 & -0.49 & -0.16\\
$V-R$ & 0.27 & 9 & 0.01 & 0.06 & 0.02\\
$V-I$ & 1.01 & 7 & 0.13(0.02) & 0.31(0.05) & 0.10(0.02)\\
\hline
\end{tabular}
\label{tab:EBV}
\end{table*} 

\citet{2004mmu..sympE...2H} finds the $(V-I)$ reddening estimate to be better behaved than the $(B-V)$ estimate, and in the case of SN~2003gd we find this also to be true. The $(B-V)$ fit yields a negative reddening which is unphysical and could be due to line blanketing as suggested by \citeauthor{2004mmu..sympE...2H}. The $(V-R)$ estimate also gives an unrealistically small reddening value, which is less than the Galactic reddening of $E(B-V)$ = 0.07 \citep{1998ApJ...500..525S}. In any case the reduced-$\chi^2$ values for the $(B-V)$ and $(V-R)$ are indicative of poor fits and were excluded from the reddening estimate. Using the $(V-I)$ colour fit we estimate the reddening to be $E(B-V) = 0.10\pm0.03$, which is comparable to that determined by \citet{2003PASP..115.1289V}. The error was determined by combining in quadrature the error from the $\chi^2$-fit and the systematic error in the reddening of SN~1999em. The error that the uncertainty in the explosion date introduces was found to be negligible.

\subsubsection{Reddening towards the neighbouring stars}

\begin{figure}
\epsfig{file = 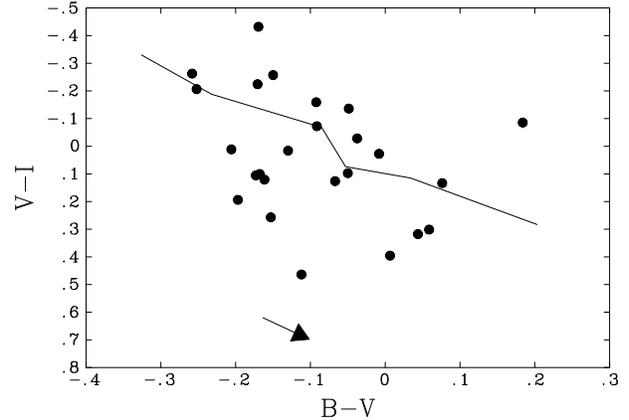, angle = -90, width = 80 mm}
\caption{A two colour diagram showing 25 stars, of M74, within 6 arcsec of the location of SN~2003gd.  The stars have been dereddened by $E(B-V)=0.13$.  The intrinsic colour sequence of supergiants \citep{2000asqu.book.....C} is shown as the solid line.   The reddening vector is indicated by the arrow.}\label{fig:red}
\end{figure}

Three colour ACS photometry (see \S\ref{sec:acs}) was used to estimate the reddening towards SN~2003gd. $(B-V)$ and $(V-I)$ colours of 25 stars within 6 arcsec of SN~2003gd were compared with the intrinsic supergiant colour sequence of \citet{2000asqu.book.....C}. The reddening was calculated using a $\chi^2$-minimisation of the displacement of the stars from the intrinsic supergiant colour sequence, for a range of values of $E(B-V)$.  The reddening vector, in the $(B-V)$/$(V-I)$ colour plane, assumed the reddening laws of \citet{1989ApJ...345..245C} with $R_V = 3.1$. Using this method the reddening was estimated as $E(B-V)=0.13\pm0.07$. The positions of the 25 dereddened stars and the intrinsic supergiant colour sequence, on the $(B-V)$/$(V-I)$ colour plane, are shown in Figure \ref{fig:red}.

\subsubsection{Reddening towards nearby H{\sc ii} regions}

The abundance gradient in M74 has been studied by \citet*{1992ApJS...78...61B}, who observed the nebular lines of 132 H{\sc ii} regions. For each H{\sc ii} region the authors calculated an extinction at H$\beta$, which gives the total extinction i. e. a combined Galactic plus extragalactic extinction. The closest H{\sc ii} regions to SN~2003gd, which were analysed, are \citeauthor{1992ApJS...78...61B} numbers 72 and 73. The logarithmic extinction $c(\rm{H}\beta)$, visual extinction $A_V$ and the colour excess $E(B-V)$ for these H{\sc ii} regions are given in Table \ref{tab:HII} along with their mean. The visual extinction was calculated from the logarithmic extinction and an extinction relationship of $A_{4861}/A_V = 1.16$ with $R_V = 3.1$ \citep{1989ApJ...345..245C}.
\begin{table}
\centering
\caption[]{Nearby H{\sc ii} regions to SN~2003gd in M74 analysed by \citet{1992ApJS...78...61B}.}
\begin{tabular}{lrrr}\hline
H{\sc ii} region & $c(\rm{H}\beta)$ & $A_V$ & $E(B-V)$\\
\hline
72 & 0.47 & 1.01 & 0.33\\
73 & 0.06 & 0.13 & 0.04\\
\hline
Mean & & & 0.19(0.15)\\
\hline
\end{tabular}
\label{tab:HII}
\end{table}

\subsubsection{Visual extinction of SN~2003gd}
A summary of the results of the methods discussed is given in Table \ref{tab:cres} along with the mean of the methods, $E(B-V) = 0.14\pm0.06$. The error on the mean was found to be unrealistically small at 0.03, therefore we have quoted the combined errors of the three methods. The Galactic extinction law with $R_V = 3.1$ was used to calculate the visual extinction, which was found to be $A_V = 0.43 \pm 0.19$. 
\begin{table}
\centering
\caption[]{Summary of the results from the three methods used in determining the reddening of SN~2003gd.}
\begin{tabular}{lr}
\hline
Method & $E(B-V)$\\
\hline
Comparison with SN~1999em & 0.10(0.03)\\ 
{\it HST} photometry & 0.13(0.07)\\
Nearby H{\sc ii} regions & 0.19(0.15)\\
\hline
Mean & 0.14(0.06)\\
\hline
\end{tabular}
\label{tab:cres}
\end{table}

\subsection{Expansion velocity of SN~2003gd}\label{sec:expvel}

\begin{figure}
\begin{center}
\epsfig{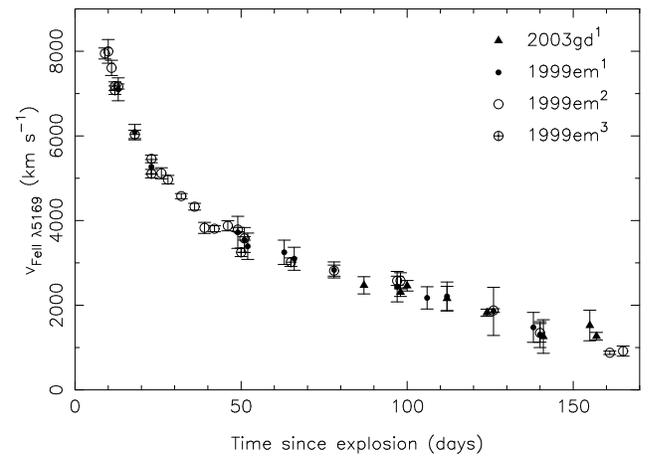}
\caption{Velocity evolution of SN~2003gd and SN~1999em derived from the minimum of the blue-shifted absorption trough of Fe {\sc ii} $\lambda$5169. The superscripts in the figure denote the source of the velocity measurements: (1) this paper, (2) \citet{2002PASP..114...35L}, (3) \citet{2001ApJ...558..615H}.} 
\label{fig:FeII99em03gd}
\end{center}
\end{figure}

We measured the expansion velocity of the ejecta of SN~2003gd from the minimum of the blue-shifted absorption trough of the Fe {\sc ii} $\lambda$5169 line, as in \citet{astro-ph/0309122}. This absorption line is asymmetric and shows signs of blending, possibly with Ti {\sc ii} and Mg {\sc i} $\lambda$5167 \citep{2003PhDT}. The absorption trough was fitted by three Gaussians using the Emission Line Fitting package ({\sc elf}) within the {\sc starlink} spectral analysis package {\sc dipso}. The minimum was found from the convolution of the three Gaussians and the error was estimated from the difference between this and a single Gaussian fit. The velocities measured in this way for SN~2003gd are listed in Table \ref{tab:FeII2003gd}. As a consistency check several epochs of SN~1999em, from \citet{2001ApJ...558..615H}, \citet{2002PASP..114...35L} and \citet{2003MNRAS.338..939E}, were also measured using the same technique. The velocities measured here are listed in Table \ref{tab:FeII1999em}. The velocity evolutions of SNe~2003gd and 1999em are plotted in Figure \ref{fig:FeII99em03gd}.
\begin{table}
\centering
\caption[]{Velocities derived from the minimum of the Fe {\sc ii} $\lambda$5169 line for SN~2003gd.}
\label{tab:FeII2003gd}
\begin{tabular}{lrrr}
\hline
Date & JD & Phase  & v$_{{\rm Fe II} \lambda 5169}$\\
& (2450000+) & (days)  & (km s$^{-1}$) \\ 
\hline
2003 Jun 14 & 2804.72 & 87  &  2514(237)\\
2003 Jun 25 & 2815.68 & 98  &  2282(126)\\
2003 Jun 27 & 2817.67 & 100 &  2398(126)\\
2003 Jul 09 & 2829.68 & 112 &  2050(126)\\
2003 Jul 21 & 2841.60 & 124 &  1934(126)\\
2003 Aug 07 & 2858.71 & 141 &  1876(872)\\
2003 Aug 22 & 2873.72 & 155 &  1586(525)\\
2003 Aug 24 & 2875.55 & 157 &  1237(126)\\
\hline
\end{tabular}
\end{table}
\begin{table}
\begin{center}
\caption[]{Velocities derived from the minimum of the Fe {\sc ii} $\lambda$5169 line for SN~1999em using the technique described here.}
\label{tab:FeII1999em}
\begin{tabular}{lrrrr}
\hline
Date & JD & Phase & v$_{{\rm Fe II} \lambda 5169}$ & Spectral\\
& (2450000+) & (days) & (km s$^{-1}$) & Source\\ 
\hline
1999 Nov 09 & 1492.14 & 13  & 7101(271) & 1\\
1999 Nov 14 & 1496.67 & 18  & 6088(182) & 2\\
1999 Nov 19 & 1501.66 & 23  & 5268(173) & 2\\
1999 Dec 15 & 1527.74 & 49  & 3722(378) & 1\\
1999 Dec 17 & 1529.74 & 51  & 3542(294) & 1\\      
1999 Dec 18 & 1530.70 & 52  & 3394(311) & 3\\
1999 Dec 29 & 1541.70 & 63  & 3250(287) & 3\\
1999 Dec 31 & 1543.66 & 66  & 3098(272) & 2\\
2000 Jan 13 & 1556.84 & 78  & 2834(189) & 1\\
2000 Feb 01 & 1575.74 & 97  & 2439(360) & 1\\      
2000 Feb 09 & 1584.35 & 106 & 2173(263) & 3\\
2000 Feb 16 & 1590.55 & 112 & 2207(341) & 3\\
2000 Mar 01 & 1604.74 & 126 & 1855(569) & 1\\     
2000 Mar 13 & 1616.51 & 138 & 1477(353) & 3\\
2000 Mar 15 & 1618.64 & 140 & 1306(305) & 1\\
\hline
\end{tabular}
\end{center}
{\footnotesize 1 \citet{2002PASP..114...35L}, 2 \citet{2001ApJ...558..615H}, 3 \citet{2003MNRAS.338..939E}}\\
\end{table}

\begin{figure}
\begin{center}
\epsfig{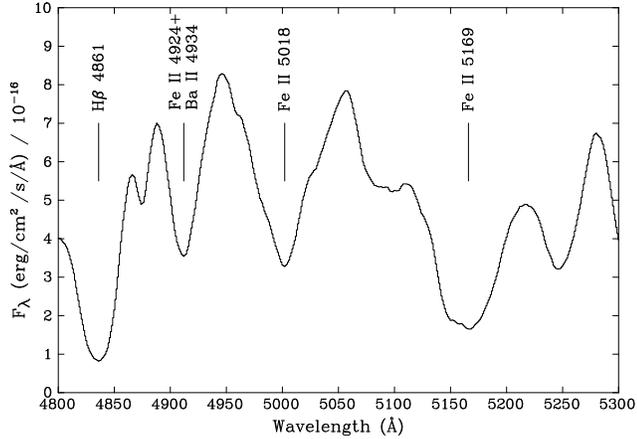}
\caption{Spectrum of SN~1999em \citep{2002PASP..114...35L} from 161 days after explosion showing blend appearing in the Fe {\sc ii} $\lambda$~5169 line.} 
\label{fig:FeII5169}
\end{center}
\end{figure}

The velocity evolution of SN~2003gd closely follows that of SN~1999em at all epochs. As the supernovae age it becomes increasingly difficult to measure the minimum of the Fe {\sc ii} $\lambda$5169 line as a blended trough begins appearing as shown in Figure \ref{fig:FeII5169}. 
\begin{figure}
\begin{center}
\epsfig{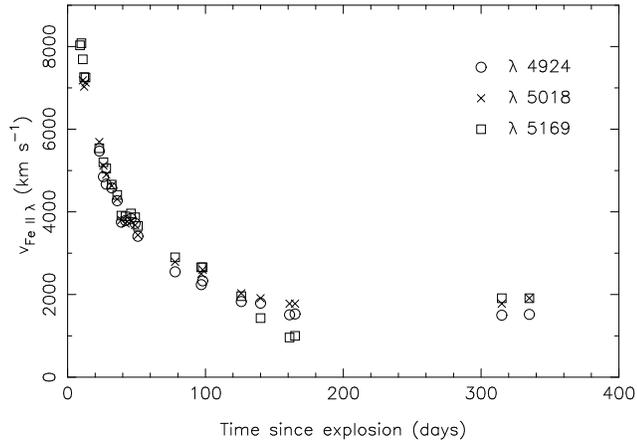}
\caption{Velocity evolution of SN~1999em derived from the minimum of the blue-shifted absorption troughs of Fe{\sc ii} $\lambda\lambda$4924, 5018, 5169, taken from \citet{2002PASP..114...35L}.} 
\label{fig:FeII1999em}
\end{center}
\end{figure}
The velocity measurements carried out in this work on the spectra of \citet{2001ApJ...558..615H}, \citet{2002PASP..114...35L} and \citet{2003MNRAS.338..939E} are comparable with the existing measurements. This demonstrates the consistency of the techniques used. Although the methods are consistent we propose that the Fe {\sc ii} $\lambda$5169 line is not a good indication of the ejecta velocity at later epochs. Figure \ref{fig:FeII1999em} shows the velocities derived from the Fe {\sc ii} $\lambda\lambda$4924, 5018, 5169 lines, taken from \citet{2002PASP..114...35L}, which are traditionally used to determine the photospheric velocity. These authors, however, point out the consistently lower velocity derived from the Fe {\sc ii} $\lambda$4924 line compared with those calculated from the other Fe {\sc ii} lines.  \citeauthor{2002PASP..114...35L} suggest that this is due to significant blending with Ba {\sc ii} $\lambda$4934, which was also found by \citet{2003PhDT}. Figure \ref{fig:FeII1999em} demonstrates this, but also shows that the Fe {\sc ii} $\lambda$5169 line, at around 140 days, gives significantly lower velocities. Although Fe {\sc ii} $\lambda$5169 yields lower velocities at these epochs, for our purposes of comparison, this is of little consequence.

\subsection{Late-time photometry and spectroscopy}\label{sec:ltspecphot}

\begin{figure}
\begin{center}
\epsfig{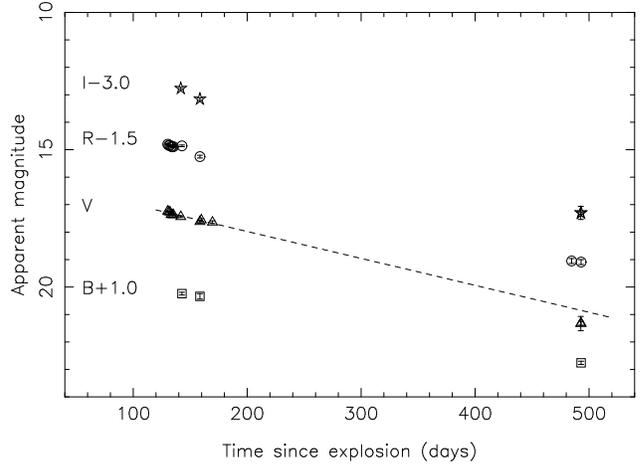}
\caption{$BVRI$ nebular and late-time photometry of SN~2003gd with a linear fit of the earlier nebular data, in the $V$-band, shown with a dashed line. The late-time photometry, in the $V$-band, is lower than what we would expect from the earlier photometry.}
\label{fig:LClt}
\end{center}
\end{figure}

\begin{figure}
\begin{center}
\epsfig{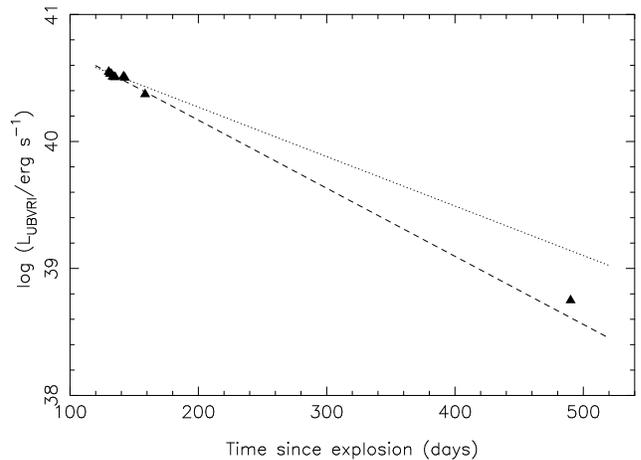}
\caption{Pseudo-bolometric (UVOIR) light curve of SN~2003gd with a linear fit of the earlier nebular data shown with a dashed line and the decay slope of $^{56}{\rm Co}$ $\rightarrow$ $^{56}{\rm Fe}$ shown with a dotted line.}
\label{fig:UVOIR03gdlt}
\end{center}
\end{figure}

\begin{figure}
\begin{center}
\epsfig{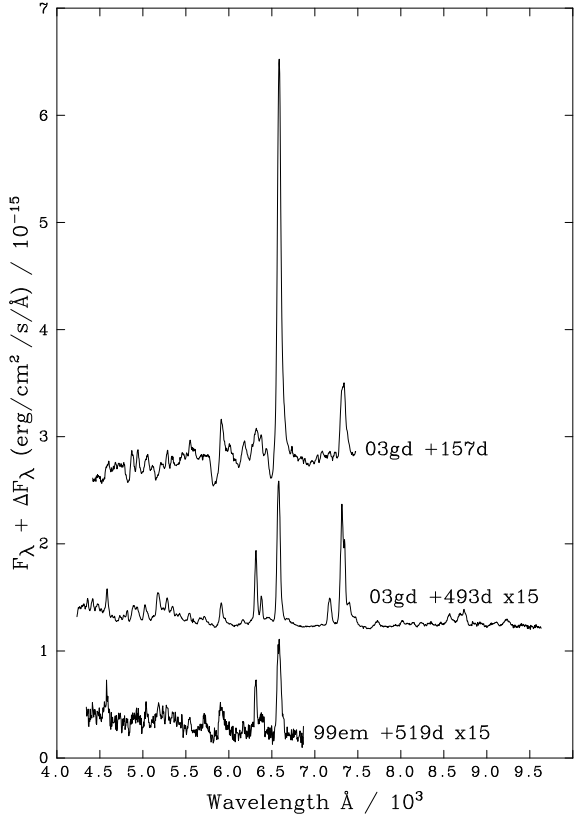}
\caption{Late-time spectrum of SN~2003gd at 493d shown with its penultimate spectrum and a spectrum of SN~1999em of a similar epoch.}
\label{fig:ltspec}
\end{center}
\end{figure}

Late-time photometry was obtained around 490 days after explosion and is plotted in Figure~\ref{fig:LClt} along with the nebular phase photometry. The details of these observations can be found in Table~\ref{tab:opphot}. A linear fit has been applied to the early nebular phase data in the $V$-band, which is shown with a dashed line. The decay rate from this fit was estimated to be 0.98~mag~100d$^{-1}$, which corresponds to the decay rate of $^{56}$Co, suggesting that little or no $\gamma$-rays escaped during this time. The pseudo-bolometric (UVOIR, see \S\ref{sec:normal?}) light curve is shown in Figure~\ref{fig:UVOIR03gdlt} where the dashed line is the fit to the early nebular data and the dotted line is the decay slope of $^{56}{\rm Co}$ $\rightarrow$ $^{56}{\rm Fe}$. A late-time spectrum was also obtained, around 493 days after explosion, and is shown in Figure~\ref{fig:ltspec} alongside the penultimate spectrum of SN~2003gd and a spectrum of SN~1999em of a similar epoch. 

\begin{figure}
\begin{center}
\epsfig{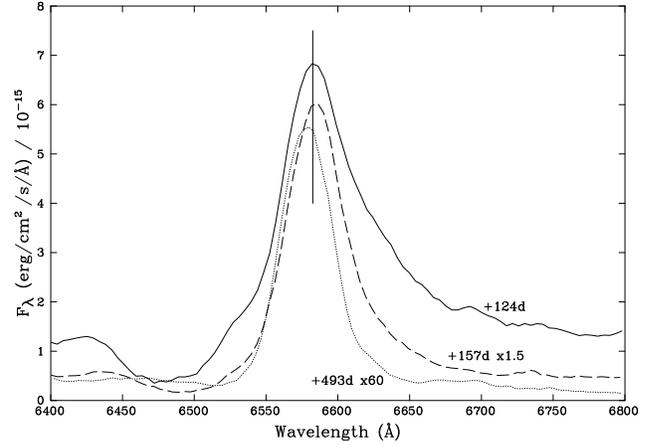}
\caption{H$\alpha$ profiles of SN~2003gd from 124d, 157d and 493d post explosion. The vertical line marks the maximum of the H$\alpha$ emission from the earliest spectrum. The peak of the H$\alpha$ emission from 157d shows a slight red-shift of the peak, but the latest spectrum clearly shows a shift to the blue.}
\label{fig:Halpha}
\end{center}
\end{figure}

Figure~\ref{fig:LClt} shows a deficit in the latest value of the $V$-band luminosity when compared with that which is expected from the linear fit. This deficit could either be due to dust formation or the supernova becoming more transparent to $\gamma$-rays, allowing leakage to occur, or a combination of both. The fit in Figure~\ref{fig:UVOIR03gdlt}, the dashed line, would suggest that the latest point is actually brighter than expected, but due to the limited data in the tail it is difficult to draw any conclusions. If we assume the $\gamma$-rays are fully thermalised, as the fit to the $V$-band data would suggest, then the luminosity is substantially less than what we would expect. \citet{2003MNRAS.338..939E} noted a blue-shift of the line profiles of [O{\sc I}] $\lambda \lambda$6300, 6364 \AA\ and H$\alpha$, and attributed this to the formation of dust. Figure~\ref{fig:Halpha} shows the profiles of H$\alpha$ from three epochs; 124d, 157d and 493d post explosion, where the vertical line marks the maximum of the H$\alpha$ emission from the earliest spectrum. The peak of the H$\alpha$ emission from 157d shows a slight red-shift, but the latest spectrum clearly shows a skewing towards the blue. The [O{\sc I}] $\lambda \lambda$6300, 6364 \AA\ lines, however, are not clear enough in the earlier spectra to verify this blue-shift. Nevertheless, the drop in luminosity at late-times, combined with the apparent blue-shift of the H$\alpha$ emission, does indeed suggest dust formed between 157 and 490 days.

\section{Distance estimates to M74}\label{sec:Dw}

\subsection{Type II-P standard candle method distance estimate}\label{sec:SCM}

\citet{1968AJ.....73.1021K} first suggested that supernovae of type Ia (SNe~Ia) could be useful distance indicators as their peak magnitudes and redshifts were found to be closely correlated. From observations of high-z supernovae it is possible to measure the history of the expansion of the universe. In recent years many separate studies have been carried out using this SNe~Ia method. One of the most recent studies, carried out by \citet{2003ApJ...594....1T}, confirmed the results of two previous studies, by \citet{1999AJ....117..707R} and \citet{1999ApJ...517..565P}, who found that the universe is actually accelerating. There is currently great debate on whether SNe~Ia are in fact standard candles and work on methods of standardising these supernovae is underway \citep[ and references therein]{2004MNRAS.349.1344A}. In order to verify the results from SNe~Ia methods, \citet{2002ApJ...566L..63H} investigated the plateau subclass of type II supernovae (SNe~II-P) for use as standard candles. Although SNe~II-P showed a wide range of luminosities at all epochs \citet{2002ApJ...566L..63H} showed that the velocity of the ejecta of SNe~II-P and their bolometric luminosities during the plateau phase are highly correlated. \citet{2002ApJ...566L..63H} formulated a Standardised Candle Method (SCM) for SNe~II-P which can be solved for the Hubble Constant, provided a suitable distance calibrator is known. Hamuy in two further papers \citep{2004mmu..sympE...2H,astro-ph/0309122} firstly confirmed the SCM using a sample of 24 SNe~II-P and then calibrated the Hubble diagram with known Cepheid distances to four of these supernovae. \citet{astro-ph/0309122} calculated the Hubble constant, using $V$ and $I$ data, to be $H_0(V) = 75\pm7$ \hub and $H_0(I) = 65\pm12$ \hub. In this paper we use the weighted average of these results, $H_0 = 72\pm6$ \hub, which is comparable to $H_0 = 71\pm2$ \hub derived as part of the {\it HST} Key Project using SNe~Ia \citep{2001ApJ...553...47F}. This result demonstrates the integrity of this method.

We use the results from \citet{astro-ph/0309122} conversely to estimate the distance to SN~2003gd, using Equations \ref{equ:V} and \ref{equ:I}:
\begin{center}
\begin{equation}
\label{equ:V}
D(V) = \frac{10^{\frac{1}{5}(V_{50}-A_V+6.249(\pm 1.35)log(v_{50}/5000)+1.464(\pm 0.15))}}{H_0}
\end{equation}
\begin{equation}
\label{equ:I}
D(I) = \frac{10^{\frac{1}{5}(I_{50}-A_I+5.445(\pm 0.91)log(v_{50}/5000)+1.923(\pm 0.11))}}{H_0}
\end{equation}
\end{center}
where $V_{50}$, $I_{50}$ and $v_{50}$ are the $V$ and $I$ magnitudes, and the expansion velocity, in km s$^{-1}$, at day 50. As SN~2003gd was in conjunction with the sun 50 days after explosion we need to infer these quantities. 

\begin{figure}
\centering
\epsfig{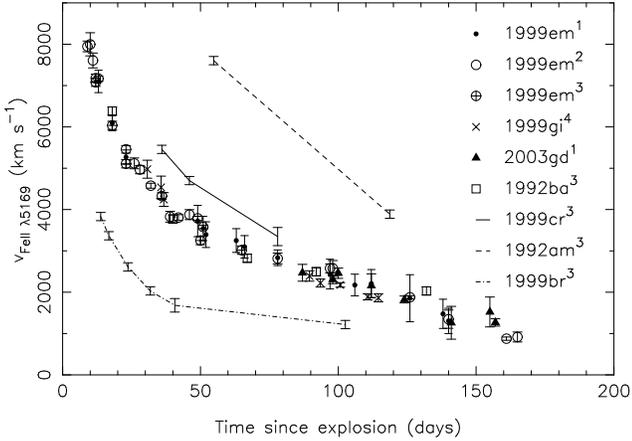}
\caption{Velocity evolution of SN~2003gd compared with other similar SNe~II-P (1999em, 1999gi and 1992ba) and contrasting SNe~II-P (1999cr, 1992am and 1999br). The superscripts in the figure denote the source of the velocity measurements: (1) this paper, (2) \citet{2002PASP..114...35L}, (3) \citet{2001ApJ...558..615H} and (4) \citet{2002AJ....124.2490L}.}
\label{fig:vcomp}
\end{figure}

The velocity evolution of SN~2003gd and SN~1999em is shown in Figure \ref{fig:FeII99em03gd}. In order to infer the velocity of SN~2003gd at day 50 from SN~1999em it is necessary to determine whether the evolution of SN~1999em is common to other similar type II-P supernovae or if it is in any way peculiar. Figure \ref{fig:vcomp} shows the expansion velocities of SN~2003gd compared to other similar SNe~II-P (SNe~1999em, 1999gi and 1992ba) and contrasting SNe~II-P (1999cr, 1992am and 1999br).
In all cases, apart from SN~1999gi, the ejecta velocities were measured from the minimum of the blue-shifted absorption trough of the Fe {\sc ii} $\lambda$5169 line. In the case of SN~1999gi the velocities were measured from the minimum of the available weak, unblended line features \citep{2002AJ....124.2490L}. SNe~1999em, 1999gi, 1992ba are all very similar SNe~II-P with comparable velocities at day 50, $v_{50}$, and towards the end of the plateau phase. The close correspondence of the velocities of these SNe with 2003gd later in the plateau phase ($\sim$100 days) suggests that SN~2003gd would also have had a similar $v_{50}$. To illustrate the difference between these and other SNe~II-P, with quite contrasting velocity evolutions, we plot SNe~1999cr, 1992am and 1999br. Within the sample of SNe~II-P studied there is no known example of a type II-P supernova which has a similar velocity evolution to that of SN~2003gd between 80-130 days which then diverges from the 1999em/1999gi/1992ba slope at day 50. This assumption allows us to estimate the velocity of SN~2003gd at day 50. Due to the similarities between SN~2003gd and SN~1999em we take the weighted average of three independent measurements of $v_{50}$ for SN~1999em and adopt this for SN~2003gd. The independent measurements are from \citet{2001ApJ...558..615H}, \citet{2002PASP..114...35L} and this paper, giving an adopted value of $v_{50} = 3694\pm981$ km s$^{-1}$ for SN~2003gd, where the error is the standard deviation of $v_{50}$ in the sample of 22 SNe~II-P from \citet{astro-ph/0309122}. The two extreme SNe, 1992am and 1999br, were excluded in this standard deviation calculation due to their peculiarities.

The apparent magnitudes for SN~2003gd at day 50 were estimated using $\Delta m$ from the $\chi^2$-fitting results (Table \ref{tab:expchi2}) and $V_{50}$ and $I_{50}$ magnitudes for SN~1999em \citep{astro-ph/0309122}, giving $V_{50} = 13.97\pm0.50$ and $I_{50} = 13.27\pm0.52$ for SN~2003gd. The errors adopted are the following errors combined in quadrature:
\begin{enumerate}
\item The error in the magnitude at day 50 for SN~1999em.
\item The error in the $\chi^2$-fit.
\item An error associated with the assumption that the light curves of SNe~2003gd and 1999em are intrinsically the same. We have estimated this from the standard deviation in the absolute magnitudes of SNe~1992ba, 1999em and 1999gi \citep{2003ApJ...582..905H}, giving $\langle M_V\rangle = -15.92\pm0.49$, $\langle M_I\rangle = -16.35\pm0.49$.
\item An error associated with the plateau length, which was taken to be 0.09 mag in $V$ and 0.17 mag in $I$.
\end{enumerate}

We use the reddening estimated in \S\ref{sec:red} to give $A_V = 0.43\pm0.19$ and $A_I = 0.26 \pm 0.11$. Using the SCM Equations (Equations \ref{equ:V} and \ref{equ:I}) and the inferred values of $v_{50}$, $V_{50}$ and $I_{50}$ for SN~2003gd discussed, we determine the distance in each band to be $D(V) = 9.52\pm4.14$ Mpc and $D(I) = 9.67\pm3.82$ Mpc. A weighted average yields a value of $D = 9.6 \pm 2.8$ Mpc. This error is statistical and comes from combining the uncertainties of each parameter in the SCM equations. \citet{astro-ph/0309122} finds a dispersion in the Hubble diagram of $0.3^{\rm{m}}$ which translates to a precision of 15\% for extragalactic distances. The distance estimate of M74 has a greater uncertainty as we have adopted large errors to adequately cover the assumptions we have made.

We could also estimate the distance of SN~2003gd by using the Cepheid distance to SN~1999em \citep[$11.71\pm0.99$ Mpc]{2003ApJ...594..247L} directly. If we assume both supernovae have the same absolute magnitudes we can use the distance modulus for NGC 1637 and apply a correction for the difference in apparent magnitudes, shown in Equation \ref{equ:mu}, to calculate the distance modulus of SN~2003gd. We use the results given in Table \ref{tab:expchi2} and the visual extinction of \S\ref{sec:red}.
\begin{center}
\begin{eqnarray}
\mu^{03gd} &=& m_0^{03gd}-M^{03gd}\nonumber\\
&=& m^{03gd}-A^{03gd}_{\lambda}-M^{03gd}\nonumber\\
&=& m^{99em}+\Delta m - {A^{03gd}_{\lambda}}-M^{03gd}\nonumber\\
&=& m_0^{99em}+A^{99em}_{\lambda}+\Delta m-{A^{03gd}_{\lambda}}-M^{99em}\nonumber\\
&=& \mu^{99em}+\Delta m+A^{99em}_{\lambda}-{A^{03gd}_{\lambda}}
\label{equ:mu}
\end{eqnarray}
\end{center}
Using this equation and the $VRI$ data we calculate $\mu(V) = 30.05\pm0.57$, $\mu(R) = 30.15\pm0.56$ and $\mu(I) = 30.16\pm0.54$. Together these give a weighted average of $\mu = 30.12\pm0.32$ or $D = 10.6\pm1.6$ Mpc which is in agreement with our SCM distance estimate.

\subsection{Brightest supergiant distance estimate using {\it HST} photometry}\label{sec:BSGs}

This method uses the correlation between the average luminosity of the brightest supergiants and the host galaxy luminosity to estimate the distance. This average luminosity should be independent of the host galaxy luminosity for there to be no distance degeneracy. This method has previously been used by \citet*{1996AJ....111.2280S} and \citet*{1996A&AS..119..499S} to estimate the distance to M74, yielding similar results. \citet{1996AJ....111.2280S} used photometry of supergiants in the disk of M74 and found $\mu = 29.3$, whereas \cite{1996A&AS..119..499S} calculated the mean modulus of the group of M74 and its dwarf companions as $\mu = 29.46$. Both of these studies used ground based imaging to determine the magnitudes and colours of the brightest stars. The resolution of these images hinders the viability of the method (as discussed in \S\ref{sec:D}), hence we have repeated this method with {\it HST} photometry of M74.

We have used the method of \citet{1996AJ....111.2280S} to determine the magnitude of the brightest supergiants. Their method firstly divides the supergiants into blue and red using their $(V-R)$ colours. The authors used $(V-R) = 0.5$ which, taking into account the foreground reddening of M74, corresponds approximately to an F8 supergiant \citep[Table 15.7]{2000asqu.book.....C}. The luminosity functions of blue and red supergiants then tell us the luminosity of the brightest supergiants, whilst removing contamination from foreground stars. This is accomplished by choosing the bin with statistically more supergiants than foreground stars and adopting this magnitude for the brightest supergiants. The number of foreground stars were calculated from field 11 of \citet{1981ApJS...47..357B}. In this work we have used an excess, of supergiants-to-foreground stars, of 2$\sigma$ to indicate a significant detection.

\begin{figure}
\begin{center}
\epsfig{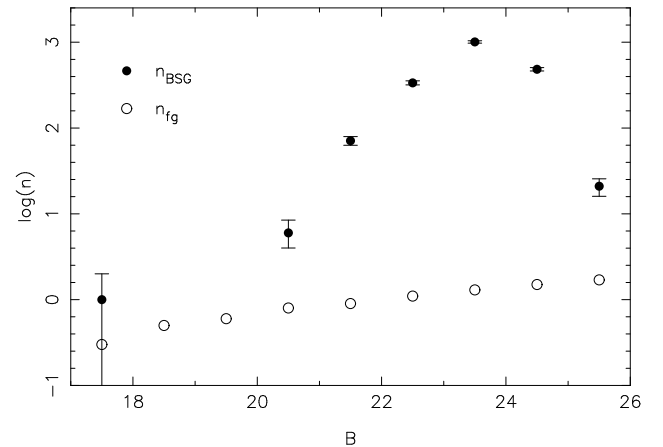}
\caption{Luminosity function in the $B$-band for BSGs where $(B-I) < 1.28$ in M74.}
\label{fig:BSGs}
\end{center}
\end{figure}

The {\it HST} (WFPC2, see \S\ref{sec:wfpc2}) photometry was firstly dereddened using the foreground extinction towards M74 from \citet{1998ApJ...500..525S} using the NED interface.\footnote{http://nedwww.ipac.caltech.edu/forms/calculator.html} The supergiants were divided into blue and red using their $(B-I)$ colours and the intrinsic colour of \citet[Table 15.7]{2000asqu.book.....C} for F8 type supergiants i.e. $(B-I) = 1.28$. We have assumed that supergiants with $(B-I) < 1.28$ are blue supergiants (BSGs) and supergiants with $B-I \ge 1.28$ are red supergiants (RSGs). The supergiants were then divided into luminosity bins and counted, where $n_{BSG}$ and $n_{RSG}$ are the number of blue and red supergiants within the bin, respectively. The resulting luminosity functions for blue and red supergiants are listed in Tables \ref{tab:BSG} and \ref{tab:ISG}, and are plotted in Figures \ref{fig:BSGs} and \ref{fig:RSGs}, in $B$ and $I$ respectively.
\begin{table}
\begin{center}
\caption[]{Luminosity functions of blue and red supergiants in $B$-band, where $n_{BSG}$ and $n_{RSG}$ are the number of blue and red supergiants, respectively, and $n_{fg}$ is the number of foreground stars predicted from \citet{1981ApJS...47..357B}, field 11.}
\begin{tabular}{lllrllrllr}\hline
\label{tab:BSG}
$B$ &&& $n_{BSG}$ &&& $n_{RSG}$ &&& $n_{fg}$\\
\hline
17-18 &&&    1(1)  &&&   0(0)  &&& 0.3\\
18-19 &&&    0(0)  &&&   1(1)  &&& 0.5\\
19-20 &&&    0(0)  &&&   0(0)  &&& 0.6\\
20-21 &&&    6(2)  &&&   3(2)  &&& 0.8\\
21-22 &&&   71(8)  &&&  10(3)  &&& 0.9\\
22-23 &&&  336(18) &&&  37(6)  &&& 1.1\\
23-24 &&& 1007(32) &&& 181(13) &&& 1.3\\
24-25 &&&  484(22) &&& 217(15) &&& 1.5\\
25-26 &&&   21(5)  &&&  72(8)  &&& 1.7\\
\hline 
\end{tabular}
\end{center}
\end{table}
\begin{table}
\centering
\caption[]{Luminosity functions of blue and red supergiants in $I$-band, where $n_{BSG}$ and $n_{RSG}$ are the number of blue and red supergiants, respectively, and $n_{fg}$ is the number of foreground stars predicted from \citet{1981ApJS...47..357B}, field 11.}
\begin{tabular}{lllrllrllr}\hline
\label{tab:ISG}
$I$ &&& $n_{BSG}$ &&& $n_{RSG}$ &&& $n_{fg}$\\
\hline 
17-18 &&&    1(1)  &&&   1(1)   &&& 1.1\\
18-19 &&&    0(0)  &&&   3(2)   &&& 1.4\\
19-20 &&&    4(2)  &&&   5(2)   &&& 1.8\\
20-21 &&&   23(5)  &&&  21(5)   &&& 2.2\\
21-22 &&&  138(12) &&& 109(10)  &&& 2.6\\
22-23 &&&  650(26) &&& 287(17)  &&& 2.8\\
23-24 &&& 1060(33) &&&  95(10)  &&& 2.8\\
24-25 &&&   43(7)  &&&   6(2)   &&& 2.8\\
25-26 &&&    7(3)  &&&   0(0)   &&& 2.8\\
\hline
\end{tabular}
\end{table}
\begin{figure}
\begin{center}
\epsfig{file = RSGlfn.epsi, angle = -90, width = 83mm}
\caption{Luminosity function in the $I$-band for RSGs where $(B-I) \ge 1.28$ in M74.}
\label{fig:RSGs}
\end{center}
\end{figure}

The number of foreground stars were calculated from the predictions of \citet{1981ApJS...47..357B} and scaled to the WFPC2 field-of-view of 5.0 sq. arcmin. A significant statistical excess of BSGs was found in the range $B = 20-21$ and RSGs in the range $I = 20-21$. The $B$ magnitude of the brightest BSGs was taken to be in the middle of this bin, $B = 20.5\pm0.5$. Using the calibrations of \citet[ Figures 10(c) and 10(f)]{1994MNRAS.271..530R} for BSGs and the apparent $B$ magnitude of M74 \citep[ B = 9.77]{1984A&A...132...20S}, the distance moduli were calculated. Since their calibrations for RSGs are only for the $K$ and $V$ bands, we converted our $I$ magnitudes to $V$ magnitudes. \citeauthor{1994MNRAS.271..530R} warn against this and point out that the brightest stars in one band will not necessarily be the brightest in another. We therefore chose the three bluest stars in the $I = 20-21$ bin. We averaged their $B-I$ colours and their $I$ magnitudes and used the intrinsic colours of \citet[ Table 15.7]{2000asqu.book.....C} to convert them to a $V$ band magnitude. We obtain $I = 20.6$ with $B-I = 1.3$ which corresponds to a F8-G0 supergiant with a $(V-I)$ colour of 0.7 and magnitude of $V = 21.4$. We then used Figures 10(a) and 10(d) of \citet{1994MNRAS.271..530R} to estimate the distance modulus for each calibration. The resulting distance moduli are listed in Table \ref{tab:SG}, where the errors are the combined errors from \citet[Table 5]{1994MNRAS.271..530R}, the error in the brightest supergiant magnitude and an error for the host galaxy reddening. We have also included an extra error of 0.5, to account for the transformation of the $I$ magnitudes.
\begin{table}
\begin{center}
\caption[]{Results of the brightest supergiant distance estimate.}
\begin{tabular}{lrrrr}\hline
\label{tab:SG}
 & 10(a) & 10(c) & 10(d) & 10(f)\\
\hline
M74 & 29.70(0.92) & 29.47(1.03) & 29.40(0.90) & 29.15(1.01)\\
\hline 
\end{tabular}
\end{center}
\end{table}
A weighted average gives $\mu = 29.44\pm0.48$ corresponding to a distance of $D = 7.7\pm1.7$ Mpc. Had we excluded the RSG result in light of the warnings of \citet{1994MNRAS.271..530R}, the distance modulus would be $\mu = 29.31\pm0.72$ or $D = 7.3\pm2.4$ Mpc. The distance we find is consistent with those of \citet[ $D = 7.2$ Mpc]{1996AJ....111.2280S} and \citet[ $D = 7.8$ Mpc]{1996A&AS..119..499S}.

It is interesting however to note that \citet{1996A&AS..119..499S} use the calibrations of \citet{1994A&A...286..718K} which are different to those of \citet{1994MNRAS.271..530R}. \citeauthor{1994A&A...286..718K} find three different calibrations, two for BSGs and one for RSGs. These are shown in Equations \ref{equ:KTV}$-$\ref{equ:KTB2}, with a standard deviation of $\sigma(M)$ = 0.30, 0.30 and 0.45 respectively.
\begin{eqnarray}
\centering
\langle M_V \rangle &=& 0.19 M_B^{gal}-4.52\label{equ:KTV}\\
\langle M_B \rangle &=& 0.35 M_B^{gal}-2.50\label{equ:KTB1}\\
\langle M_B \rangle &=& -0.51(B-B^{gal})-4.14\label{equ:KTB2}
\end{eqnarray}
The distance modulus can be calculated from Equations \ref{equ:KTV} and \ref{equ:KTB1} using an iterative approach starting with the absolute magnitude from \citet[ $M_B^{gal} = -20.23$]{1984A&A...132...20S}. Using this method we found the distance moduli to be $\mu(V) = 29.71 \pm 0.77$ and $\mu(B) = 30.12 \pm 0.59$, where the errors are the combined errors of the relationship, the error in the magnitude of the brightest supergiants and the reddening of the galaxy, which were both discussed previously. \citet{1996A&AS..119..499S} used only the calibration given in Equation \ref{equ:KTB2} for BSGs, which does not suffer from distance degeneracy. The authors found only one star red enough to be a RSG, but concluded it was a foreground star due to its magnitude. The distance modulus calculated from this relationship is $\mu(B) = 30.11 \pm 0.67$. A mean using the first two relationships only, gives a distance modulus of $\mu = 29.91 \pm 0.49$ corresponding to a distance of $D = 9.6 \pm 2.2$ Mpc. This is a difference of 1.9 Mpc using the same data, but with two different calibrations.

\subsection{Summary and discussion of distance estimates}\label{sec:D}

\begin{table}
\begin{center}
\caption[]{Summary of distance estimates.}
\begin{tabular}{lrrr}\hline
\label{tab:D}
Method & Source & Distance & Mean\\
& & (Mpc)& \\
\hline
SCM                   & 1 & 9.6(2.8) & 9.6(2.8)\\
\\
Brightest supergiants & 1 & 7.7(1.7) & 8.1(2.3)\\
                      & 1 & 9.6(2.2) &\\
                      & 2 & 7.2(2.0) &\\
                      & 3 & 7.8(1.6) &\\
\\
Kinematic             & 4 & 10.1 & 10.2(2.6)\\
                      & 5 & 11.1 & \\
                      & 6,7 & 9.5 &\\
\hline
Mean & & & 9.3(1.8)\\
\hline
\end{tabular}
\end{center}
1~this paper,
2~\citet{1996AJ....111.2280S},
3~\citet{1996A&AS..119..499S},
4~\citet{1988ngc..book.....T},
5~\citet{1998AJ....116..673F},
6~\citet{1998A&AS..130..333T},
7~http://leda.univ-lyon1.fr
\end{table}
A summary of distances to M74 from three different methods is listed in Table \ref{tab:D}. As well our SCM and brightest supergiant distance estimates, derived in \S\S\ref{sec:SCM} and \ref{sec:BSGs} respectively, we have collated three kinematic and two other brightest supergiant distance estimates from the literature. The kinematic distances shown in the table, use the heliocentric velocity of 656 km s$^{-1}$ and apply three different models of infall onto the Virgo cluster. The values have been corrected for a Hubble constant of $H_0 = 72$~km~s$^{-1}$~Mpc$^{-1}$. The simple mean of these distances is also listed in the table, where the error is a combination of the error on the mean and the uncertainty due to the cosmic thermal velocity dispersion of 187 km s$^{-1}$ \citep{2000ApJ...530..625T}. The accuracy of these kinematic distances is limited not only by the observed velocity dispersion around the Hubble Flow, but on an accurate model to account for the infall onto Virgo and the Great Attractor. Depending on the model the values range from $9.5-11.1$ Mpc, which is a 17\% difference, hence there is a large uncertainty associated with this method. 

The brightest supergiants distances listed in Table \ref{tab:D} come from three different sources. The first two distances, $D = 7.7\pm$1.7 and $9.6\pm2.2$~Mpc, are derived in \S\ref{sec:BSGs}. The third, $D = 7.2\pm2.0$~Mpc, is the result of \citet{1996AJ....111.2280S} where the error is a combination of the error in the size of the authors' bin and the error associated with the calibration used \citep[ Figure 10a]{1994MNRAS.271..530R}. The fourth, $D = 7.8\pm1.6$~Mpc, is from \citet{1996A&AS..119..499S} where the error is the combined error associated with the relationship used \citep{1994A&A...286..718K} and the magnitude of the brightest BSGs. The simple mean is also listed in Table \ref{tab:D} where the error is a combination of the statistical error and the systematic error of $\Delta \mu = 0.59$ which corresponds to a distance of 2.2 Mpc \citep{1994MNRAS.271..530R,1994A&A...286..718K}. Although the error in the average distance is the lowest of the three methods, the estimates range from $7.2-9.6$ Mpc which is a 26\% difference. There are large systematic uncertainties associated with this method and the derived distance also appears to be calibration dependent.

\citet{1994MNRAS.271..530R} re-examine most of the existing observational data and investigate claims that the luminosity of the brightest BSGs exhibit a dependence on parent galaxy luminosity, introducing a distance degeneracy. Not only do they confirm this, a similar dependence is also found for the brightest RSGs, which was previously thought to be independent of galaxy luminosity. \citeauthor{1994MNRAS.271..530R} also find that the resulting error in the distance modulus from this method is far greater than previously believed. Whatever calibration they used, the error in the distance modulus was found never to be less than $0.55$~mag. \citet*{1992MmSAI..63..465P} and \citet{1994A&A...286..718K} also find such dependencies but find a smaller error in the regression of $0.30$ mag. \citet{1994MNRAS.271..530R} conclude that this method should not be used in determining the distance to single galaxies. Besides these calibration problems, this method also suffers from crowding, misidentification of supergiants with H{\sc ii} regions, foreground stars, unresolved clusters and OB associations. These problems associated with poor resolution cause the distance to be systematically underestimated. This is illustrated by \cite{2003ApJ...594..247L} who derive a Cepheid distance to NGC 1637 that is 50\% larger than the brightest RSG distance estimate of \citet{1998AJ....115..130S}.

The estimates from this paper use {\it HST} photometry which has much better resolution than ground based observations and should resolve individual stars from H{\sc ii} regions and compact clusters, reducing this systematic error. If this systematic error is present we should find larger distances, which is what is found using the calibrations of \citet{1994A&A...286..718K}. Although our first estimate is comparable to both \citet{1996A&AS..119..499S} and \citet{1996AJ....111.2280S} our second estimate using a different relationship differs by 2~Mpc. It is worthy of note that whereas we have used all four of the calibrations of \citet{1994MNRAS.271..530R}, \citet{1996AJ....111.2280S} have used just that of Figure 10(a) for RSGs. If we look at the individual estimates of $\mu$ in Table~\ref{tab:SG} we see that $\mu(a) = 29.70$ is consistent with our results using \citeauthor{1994A&A...286..718K}, whereas our other results, $\mu(c$$\rightarrow$$f)$, are not. If we use the parameters employed by \citet{1996A&AS..119..499S} to calculate the distance modulus of M74, using the alternative relationships of \citet[Figure 10c]{1994MNRAS.271..530R}, we find $\mu = 28.92$ instead of their $\mu = 29.3$, which is a difference of 1.2 Mpc. Table \ref{tab:comp} shows the coefficients of the relationship $M = aM_B^{gal}+b$, found by each of the two different sources plus a third from \citet{1992MmSAI..63..465P}. The coefficients of the relationship for RSGs, $M(V)$, are consistent, but when we consider those of the BSGs relation, $M(B)$, there is a disagreement between that of \citet{1994MNRAS.271..530R} and the other two papers. \citeauthor{1994MNRAS.271..530R} find a much shallower relation resulting in the discrepancy between distance moduli. Given the agreement between the coefficients we consider the larger distance modulus to be more realistic.

\begin{table}
\begin{center}
\caption[]{Summary of coefficients from \citet{1992MmSAI..63..465P}, \citet{1994A&A...286..718K} and \citet{1994MNRAS.271..530R} for $M = aM_B^{gal}+b$.}
\begin{tabular}{lrrrr}\hline
\label{tab:comp}
$M$ & a & b & $\sigma (M)$ & Source\\
\hline
$M(V)$ & 0.21 & -4.01 & $\gtrsim$0.3 & 1\\
& 0.19 & -4.52 & 0.30 & 2\\
& 0.21 & -4.10 & 0.58 & 3\\
$M(B)$ & 0.36 & -2.29 & $\sim$0.56 & 1\\ 
& 0.35 & -2.50 & 0.30 & 2\\
& 0.28 & -3.45 & 0.90 & 3\\
\hline
\end{tabular}
\end{center}
1~\citet{1992MmSAI..63..465P},
2~\citet{1994A&A...286..718K},
3~\citet{1994MNRAS.271..530R}
\end{table}

The SCM distance estimate also has its uncertainties. This method requires data at a phase of 50 days which is not always available. In order to obtain the necessary photometric data, \citet{2004mmu..sympE...2H} interpolated the observed $V$ and $I$ magnitudes to 50 days post explosion. In most cases the author was also able to interpolate the velocities by fitting a power-law to the observed velocities \citep{2001PhDT}, to an accuracy of $\pm 300$ km~s$^{-1}$. In four cases though it was necessary to extrapolate, introducing an error of 2000~km~s$^{-1}$. The redshifts used in \citet{2004mmu..sympE...2H} were derived from the observed heliocentric redshifts and converted to the Cosmic Microwave Background frame in order to remove their peculiar motion. This in itself is limited by the accuracy of the model and by the cosmic thermal velocity. \citeauthor{2004mmu..sympE...2H} noted that the least satisfactory aspect of the SCM is the dereddening method based on the comparison of supernova colour curves, which yields different results depending on what colour is used. It also depends on the accuracy of the reddening determination of the model supernova. Whilst not as accurate as distances derived from SN~Ia, the SCM does give distances to within 15\%. This method also gives comparable values of H$_0$ to those of \citet{2001ApJ...553...47F}.

In \S\ref{sec:expvel} we highlighted the difficulties in measuring the minimum of the blue-shifted trough of the Fe {\sc ii} $\lambda$5169 line at later epochs. We found this was not as much of an issue at earlier epochs where the velocity is needed for the SCM. A large uncertainty however was introduced in our application of the SCM, due to our lack of data at 50 days. Although this is the case we have been conservative in our error analysis and the SCM distance of $D = 9.6\pm2.8$ Mpc is consistent with other distance estimates and associated errors.

Each of the three different methods to measure distance have similar uncertainties and all have shortcomings. We have therefore taken an unweighted mean of the results of each method to give an absolute distance of $D = 9.3\pm1.8$~Mpc, where the error is the error on the mean.

\section{Estimate of ejected $^{56}$Ni mass}\label{sec:Ni}
The comparison of the light curve of SN~2003gd with that of SN~1999em in Figure \ref{fig:03gd99emLC}, highlighted a major difference between the two supernovae. The lower tail luminosity was attributed to a smaller amount of nickel synthesised in the explosion as was also suggested by \citet{2003PASP..115.1289V}. Here we use three different methods to ascertain the mass of $^{56}$Ni produced. It was first of all estimated using the method of \citet{2003ApJ...582..905H}, secondly using a direct comparison with the light curve of SN~1987A and lastly using the new though unconfirmed method of \citet{2003A&A.404..1077E}.

\subsection{Nickel mass from bolometric luminosity of exponential tail}

\citet{2003ApJ...582..905H} derives $M_{\rm{Ni}}$ from the bolometric luminosity of the exponential tail, assuming that all of the gamma rays due to $^{56}\rm{Co}$ $\rightarrow$ $^{56}{\rm Fe}$ are fully thermalised. We first convert our $V$-band photometry in the tail to bolometric luminosities using the formula given in Equation 1 of \citet{2003ApJ...582..905H} which is given here in Equation \ref{equ:bol}. The bolometric correction is $BC = 0.26\pm0.06$ and the additive constant converts from Vega magnitudes to cgs units \citep{2001PhDT,2003ApJ...582..905H}. We have used the distance derived in \S\ref{sec:D}.
\begin{equation}
\log \left(\frac{L}{{\rm erg\:s}^{-1}}\right) = \frac{-(V-A_V+ BC)+5\log D-8.14}{2.5}\label{equ:bol}
\end{equation}
\begin{equation}
M_{{\rm Ni}}= 7.866\times10^{-44} L \exp \left[ \frac{(t-t_0)/(1+z)-\tau_{{\rm Ni}}} {\tau_{\rm Co}} \right] {\rm M}_{\odot} \label{equ:MNi}
\end{equation}
The nickel mass was then found using Equation 2 of \citet{2003ApJ...582..905H} given here in Equation \ref{equ:MNi}, where $t_0$ is the explosion epoch, $\tau_{{\rm Ni}} = 6.1$ days is the half-life of $^{56}$Ni and $\tau_{{\rm Co}} = 111.26$ days is the half-life of $^{56}$Co. Using this method we estimated $M_{\rm Ni}$ for each of the points in the tail and by taking a simple average of these we found $M_{\rm Ni} = 0.016^{+0.019}_{-0.008}$ \msun. This is approximately a third of that produced in SN~1999em, hence the lower tail luminosity. There is a large uncertainty associated with this value as the nickel mass calculation is strongly dependent on the explosion time, extinction and the distance, all of which are not well known. We have therefore calculated the range of nickel masses, for each point, from the extreme values for each parameter in Equations \ref{equ:bol} and \ref{equ:MNi}. We have taken an average of the maximum values and minimum values and have estimated a more appropriate error from these.

\subsection{Nickel mass from a direct comparison to SN~1987A light curve} 

The $^{56}$Ni mass was also estimated from the difference in the UVOIR light curves of SN~2003gd and SN~1987A, assuming the same $\gamma$-ray deposition. A $\chi^2$-fitting algorithm was used to shift the light curve of SN~1987A onto that of SN~2003gd to find the best fit. When constructing the light curve of SN~1987A a distance of 50~kpc was adopted. The difference in log luminosity was found to be $\log (L^{87A}/L) = 0.664 \pm 0.193$. Equation~\ref{equ:Ni87A} was then used to scale the nickel mass of SN~1987A, which was taken to be 0.075 \msun\ \citep[e.g.][]{1998ApJ...498L.129T}, to estimate that of SN~2003gd.
\begin{equation}
M_{\rm Ni} = 0.075 \times \left(\frac{L}{L^{87A}}\right) {\rm M}_{\odot}
\label{equ:Ni87A}
\end{equation}
Using this method we find $M_{\rm Ni} = 0.016 \pm 0.008$ \msun\ where the error is the error associated with the fit combined with an error of 0.003 \msun\ related to the uncertainty in the explosion date. This method is however still dependent on the distance to the supernova although the error in the nickel mass amply accounts for this.

\subsection{Nickel mass from ``steepness of decline'' correlation} 

\citet{2003A&A.404..1077E} reported a correlation between the rate of decline in the V-band, from the plateau to the tail, and the nickel mass estimated from the SN~1987A method. The authors defined a ``steepness'' parameter, S, which is the maximum gradient during the transition in mag d$^{-1}$. A sample of ten SNe~II-P were used in determining the best linear fit which is given in Equation \ref{equ:S}.
\begin{equation}
\log \left(\frac{M_{\rm Ni}}{{\rm M}_{\odot}}\right) = -6.2295 S - 0.8147
\label{equ:S}
\end{equation}
The ``steepness'' parameter for SN~2003gd was found to be 0.25 which yields the rather low result of $M_{\rm Ni} = 0.004^{+0.002}_{-0.001}$ \msun. The error was estimated from the scatter around the linear fit in the $\log (M_{\rm Ni}/$\msun$)/S$ plane, $\log (M_{\rm Ni}/$\msun$) = 0.147$, and Equation \ref{equ:S} was used to translate this into a mass. This error is probably much larger due to the error in the fit.

\subsection{Discussion of nickel mass estimate}\label{sec:concni}

The three independent methods, to obtain nickel mass, give two concurrent results of 0.016 \msun\ and one rather low estimate of 0.004 \msun. The first two methods are both dependent on the explosion date and the distance, although the errors sufficiently accommodate these. The method of \citet{2003ApJ...582..905H} assumes that the $\gamma$-rays are fully thermalised, but as the slope of the exponential tail of SN~2003gd follows the decay of $^{56}{\rm Co}$ this is not unreasonable. The direct comparison to the light curve of SN~1987A assumes that SN~2003gd deposited a similar fraction of $\gamma$-rays as SN~1987A. As the late-time decline of SN~1987A was very close to the decay rate of $^{56}$Co, this assumption is also not unfounded. The method of \citet{2003A&A.404..1077E} has the advantage of being independent of explosion epoch and distance although it is an unconfirmed method. The correlation itself is distance dependent and relies on the SN~1987A method. The result is very low and would suggest that the adopted distance of 9.3 Mpc is out by a factor of two, which is unlikely. Given that the first two results are concurrent, and the errors adequately reflect the uncertainties, we take the average of the first two results and estimate a nickel mass of $M_{\rm Ni} = 0.016^{+0.010}_{-0.006}$ \msun.

\section{Model comparison}\label{sec:zamp}

We compared the data of SN~2003gd with the output of a
semi-analytic light curve code \citep[see][ for a
detailed description of the code and the adopted initial
conditions]{2003MNRAS.338..711Z}. The goal of this comparison is twofold. First of all,
we want to put some constraints on the input parameters (in
particular the opacity) of the semi-analytic code by checking to see whether
the mass estimate from the code is consistent with the observed
measurement of the progenitor mass of $8^{+4}_{-2}$ \msun\ \citep{2004Sci...303..499S}. Secondly, we would also like to
use the code to derive some limits on the SN distance and explosion date, which are rather uncertain. The code performs an estimation of the ejected
envelope parameters from a simultaneous comparison of the observed
and computed light curve, photospheric gas velocity and continuum
temperature. The main input parameters of the envelope are listed in
Table \ref{tab:zamp}. The fraction of the initial energy that goes into
kinetic energy and the colour correction factor (that measures the
deviation of the continuum radiation temperature $T_c$ from the
blackbody effective temperature $T_{eff}$) are input constants and
are fixed at the values 0.5 and 1.2, respectively. An important
physical quantity that must be specified and that may affect in a
significant way theoretical mass estimates is the envelope gas
opacity $\kappa$.

The parameters of the model fits, corresponding to different choices of
the explosion epoch and distance, are reported in Table \ref{tab:zamp}, where the estimated error in the parameters is $\sim$20-30\%. For
models A and B we assume that the phase identification derived in
\S\ref{sec:epoch} is correct, implying that the explosion epoch is JD 2452717
(86 days before discovery). The two models differ in the adopted
SN distance, $D=7.3$ Mpc for model A and $D=9.2$ Mpc for model B.
Model C assumes JD 2452737 ($\sim$65 days before discovery) for
the explosion epoch and $D=7.3$ Mpc for the distance.

\begin{figure}
\begin{center}
\epsfig{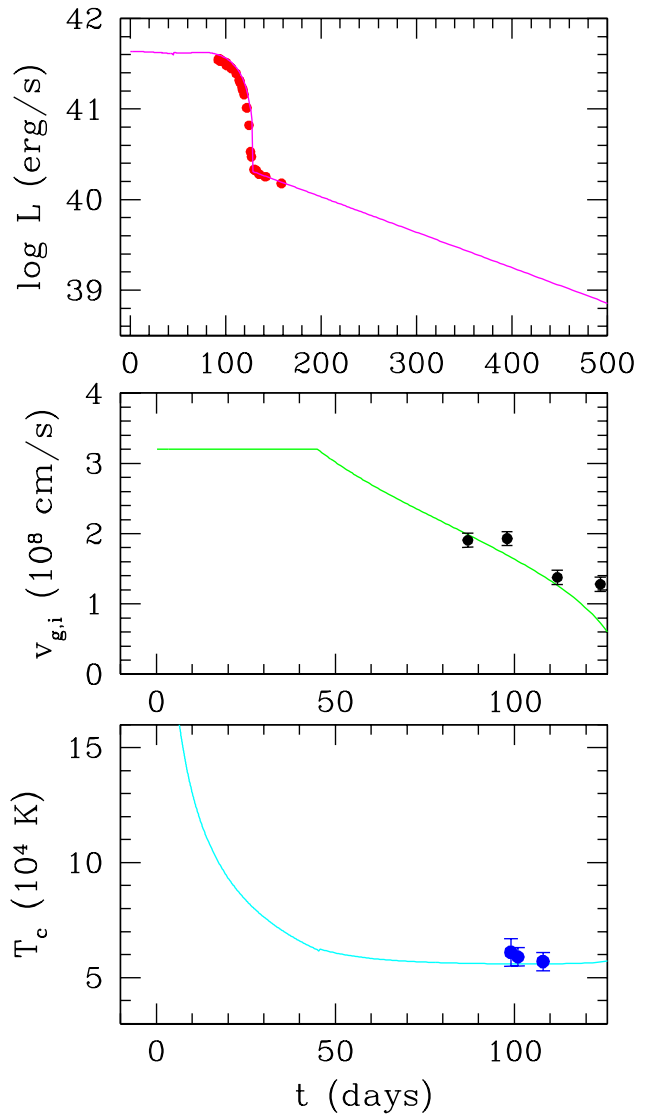}
\caption{Results for model A showing $V$-band luminosity, $L$, velocity of metal (Sc {\sc ii} $\lambda$ 6246) lines, $v_{g,i}$, and continuum temperature, $T_c$, as a function of time
for SN~2003gd. The adopted distance modulus and the estimated total
reddening are $\mu = 29.32$ and $A_V = 0.42$. The assumed explosion date
is JD 2452717. The solid lines represent the curves from the semi-analytic
model of \citet{2003MNRAS.338..711Z}.}
\label{fig:zampm}
\end{center}
\end{figure}

The results for model A are shown in Figure \ref{fig:zampm}. The overall
agreement with the observed light curve, photospheric gas velocity
and continuum temperature is rather good. The data are reproduced
by a slightly under-energetic explosion ($E=1.4\times10^{51}$ erg) that has
ejected $M=11$ \msun\ and $M_{\rm Ni}=9\times 10^{-3}$ \msun. The comparison of the inferred value of $M$ and the
observed measurement of the progenitor mass shows that the star
was probably in the upper end of the estimated mass range and sets
an interesting constraint on the average opacity of the ejected
envelope gas. In fact, values of $\kappa \le 0.3$ cm$^2$ g$^{-1}$
result in ejected envelope masses in excess of 11 \msun, barely consistent with the progenitor mass estimated from direct pre-explosion observations, once the compact remnant mass is taken into account \citep{2004Sci...303..499S}. Thus, large opacities are needed, in the range $\kappa=0.3-0.4$ cm$^2$ g$^{-1}$ ($\kappa=0.34$ cm$^2$ g$^{-1}$ in Figure \ref{fig:zampm}). Opacities in this range are appropriate for an envelope comprised mostly of hydrogen, with a composition close to solar. This suggests that little mixing has occurred in SN~2003gd, the helium and other elements synthesised during nucleosynthesis remaining buried beneath the envelope.

Model B tests the agreement between the model and the data assuming a different SN distance. This model has a rather typical explosion energy ($E=1.8\times10^{51}$ erg), while the ejected Ni mass ($M_{\rm Ni}=1.5\times 10^{-2}$ \msun) is about 5 times lower than the ``average'' value for a normal type II-P SN ($M_{\rm Ni}=7.5\times 10^{-2}$ \msun). We note, however, that there is evidence to suggest a continuous distribution of $M_{\rm Ni}$ in type II-P SNe (see Figure \ref{fig:hamuy}). The ejected envelope mass in this model is $M=13$ \msun. Considering the mass of the compact remnant, this result indicates that $M$ may be too large to be consistent with a progenitor of $\le 12$ \msun. Therefore, if the phase of the SN at discovery is $\sim$90 days, the distance should be probably lower than 9 Mpc, in the lower end of the estimated range of distances.

The uncertainty in the explosion epoch, however, is rather large and 
it cannot be ruled out that the discovery epoch is actually
closer to the explosion date than estimated. Model C was run to
test this possibility. We find that the data of SN~2003gd are
reproduced by a fairly low energy explosion ($E=0.7\times10^{51}$ erg) with the
ejection of 6 \msun\ and a small amount of Ni ($M_{\rm
Ni}=7\times 10^{-3}$ \msun). In this case smaller values of
the opacity and/or larger distances would make the model still in
agreement with the observations. However, an event of this type
would have a plateau lasting $\le 90$ days, which is shorter than
what is commonly observed in type II-P SNe. Furthermore, the
theoretical photospheric gas velocity appears to overestimate
significantly ($>$20\%) the observed line velocity at discovery.

\begin{table*}
\caption{Theoretical parameters of SN~2003gd from the semi-analytic model of \citet{2003MNRAS.338..711Z}}
\begin{minipage}{\linewidth}
\begin{center}
\begin{tabular}{lrrrrrrrr}
\hline
 & $R_0$ & $M$ & $M_{\rm Ni}$ & $V_0$ & $E$ &
 $\kappa$ & $t_{rec,0}$ & $T_{eff}$ \\
 & ($10^{13}$ cm) & (\msun) & (\msun) & ($10^8$ cm s$^{-1}$) &
 ($10^{51}$ erg) & (cm$^2$ g$^{-1}$) & (days) & (K) \\
\hline
model A   & 2.3 & 11 & $9\times 10^{-3}$ & 3.2
          & 1.4 & 0.34 & 45 & 4600 \\
model B   & 2.7 & 13 & $1.5\times 10^{-2}$ & 3.4
          & 1.8 & 0.34 & 40 & 5000 \\
model C   & 2   & 6  & $7\times 10^{-3}$ & 3.2
          & 0.7 & 0.34 & 45 & 4600 \\
\hline
\label{tab:zamp}
\end{tabular}
\end{center}
\begin{tabular}{ll}
$R_0$ initial radius of the ejected envelope & $E$ initial thermal+kinetic energy of the ejecta\\      
$M$ ejected envelope mass & $\kappa$ gas opacity\\
$M_{\rm Ni}$ ejected $^{56}$Ni mass & $t_{rec,0}$ time when the envelope starts to recombine\\
$V_0$ velocity of the envelope at the outer shell & $T_{eff}$ effective temperature during recombination\\
\end{tabular}
\end{minipage}
\end{table*}

\section{Discussion}\label{sec:diss}

\subsection{SN~2003gd - a typical type II-P supernova?}\label{sec:normal?}


Unfortunately SN~2003gd was discovered $\sim$86 days after explosion (\S\ref{sec:epoch}) when it appeared from behind the sun. Since valuable information about this supernova was lost we have made various assumptions about the supernova based on the striking spectroscopic similarities with SN~1999em and their comparable progenitor masses. Although SN~2003gd is spectroscopically very similar to SN~1999em it is fainter in its tail luminosity due to the lower nickel mass synthesised in the explosion. \citet{2003ApJ...582..905H} presented evidence to suggest that the plateau luminosities and velocities are correlated with the amount of nickel produced (see also Figure \ref{fig:hamuy}). The lower tail luminosity may therefore suggest that SN~2003gd was intrinsically much fainter than SN~1999em, although the spectral similarities and velocities at the end of the plateau phase would suggest otherwise. A number of papers have recently featured peculiar low-luminosity, low-nickel type II-P SNe \citep[e.g.][]{2003MNRAS.338..711Z,astro-ph/0310056,2004MNRAS.347...74P}, but the question is, could SN~2003gd also belong to that group and if so, were our earlier assumptions justified? 

\begin{figure}
\begin{center}
\epsfig{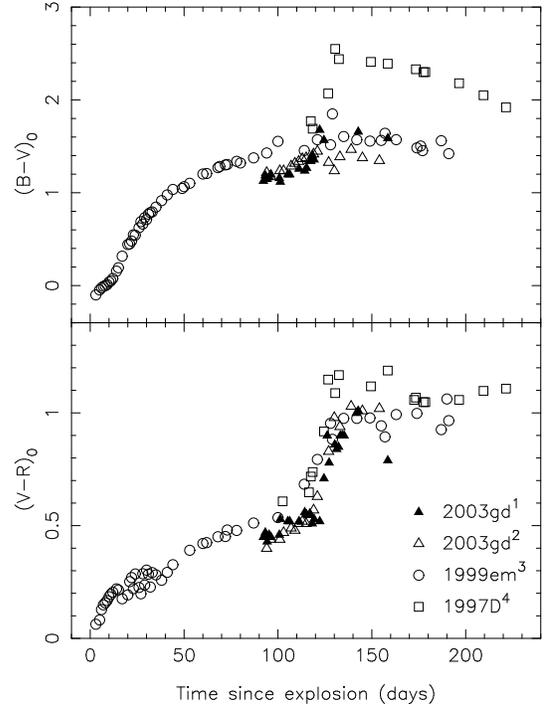}
\caption{Intrinsic $(B-V)$ and $(V-R)$ colour curves of SN~2003gd alongside those of the prototypical peculiar ``faint'' SN~1997D and the ``normal'' SN~1999em. The superscripts in the figure denote the source of the photometry: (1) this paper, (2) \citet{2003PASP..115.1289V}, (3) \citet{2001ApJ...558..615H} and (4) \citet{2001MNRAS.322..361B}.}
\label{fig:diffSNc}
\end{center}
\end{figure}

\citet{2003A&A.404..1077E} and \citet{2004MNRAS.347...74P} noted a rapid excess in the $(B-V)$ and $(V-R)$ colours of the low-luminosity, or ``faint'', SN~1997D at the end of the photospheric phase (see Figure~\ref{fig:diffSNc}). This behaviour was also observed by \citet{2004MNRAS.347...74P} in SN~1999eu, another ``faint'' supernova, in the form of a sharp spike. \citeauthor{2004MNRAS.347...74P} suggested that the excess in colour may be characteristic of this low-luminosity sub-group. Figure~\ref{fig:diffSNc} shows the intrinsic colour curves of SN~2003gd alongside those of the prototypical peculiar ``faint'' SN~1997D \citep{1998ApJ...498L.129T,2001MNRAS.322..361B,astro-ph/0310056} and the model ``normal'' SN~1999em. The $(B-V)_0$ colour curve of SN~2003gd does rise more rapidly than that of SN~1997D, but it does not reach a significant excess. The SN~2003gd colours of \citet{2003PASP..115.1289V} also do not display this steeper rise. The $(V-R)_0$ colour evolution of SN~2003gd is slightly more rapid than that of SN~1999em, at the end of the plateau phase, but it is nearer to the evolution of the ``normal'' supernova. There may be a small spike in the colour of SN~2003gd at the end of the plateau phase at around 120 days, but the photometry is sparse and it is therefore difficult to confirm its existence. We would also note that there may also be a slight peak in the colour curve of SN~1999em at $\sim$8 days later. From these colour evolutions SN~2003gd appears to be more like a ``normal'' type II-P supernova. In order to further investigate our assumptions we compared the UVOIR light curves and spectra of SN~2003gd with examples of ``normal'' and ``faint'' type II-P supernovae.

\begin{figure}
\begin{center}
\epsfig{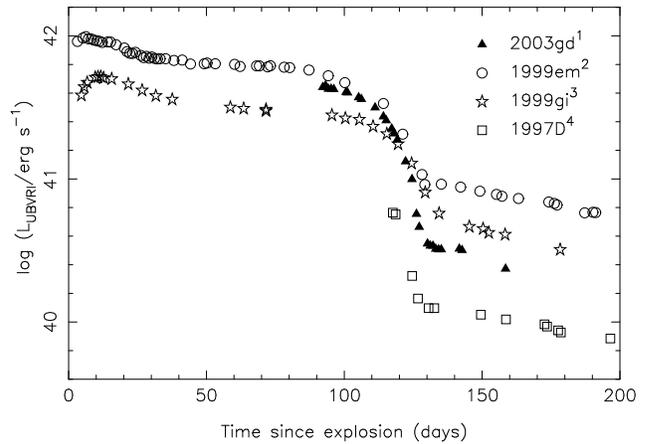}
\caption{Comparison between the $L_{\rm UBVRI}$ light curves of SNe~2003gd, 1999em, 1999gi and 1997D which have nickel masses of 0.016, 0.048, 0.022 and 0.006 \msun, respectively. The nickel masses for SNe~1999em and 1999gi are from \citet{2003ApJ...582..905H}, and the nickel mass for 1997D is from \citet{2004MNRAS.347...74P}, which are corrected for the revised distances used here. The superscripts in the figure denote the source of the photometry: (1) this paper, (2) \citet{2001ApJ...558..615H}, (3) \citet{2002AJ....124.2490L} and (4) \citet{2001MNRAS.322..361B}.}
\label{fig:UVOIR}
\end{center}
\end{figure}
The UVOIR light curve of SN~2003gd is shown in Figure~\ref{fig:UVOIR} along with those of three other type II-P supernovae, SNe~1999em, 1999gi and 1997D where SN~1999gi is a ``normal'' plateau event. These UVOIR light curves were calculated by summing up all the contributions from the $UBVRI$ observations of each supernova. The few observations in $U$ were used to estimate the contribution to the total flux from that band. In the case of SN~1999gi, where there was no $U$-band photometry, the $U$-band contribution was taken to be the same as that for SN~1999em. The distances used to calculate the UVOIR light curves were $11.4\pm 0.9$, $10.0\pm 0.8$ and $15.3\pm 1.9$ Mpc for SNe~1999em, 1999gi and 1997D respectively. The distance used for SN~2003gd is that discussed in \S\ref{sec:D}. The distance of SN~1999em used here is the straight average of those discussed by \citet[ Table 9]{2003ApJ...594..247L}. The distance of SN~1999gi, who's host galaxy was NGC 3184, was taken to be the simple mean of the distance estimates found in the literature, which are summarised in Table \ref{tab:99gi} where the error is the error on the mean.  A literature search for the host galaxy of SN~1997D, NGC 1536, returned two kinematic distances, 14.0 and 16.7 Mpc, from \citet{1988ngc..book.....T} and LEDA, which yield an average of $15.3\pm 1.3$ Mpc. All kinematic distances were corrected for a Hubble Constant of $H_0 = 72 $ km s$^{-1}$ Mpc$^{-1}$. The plateau luminosity of SN~2003gd is similar to that of SN~1999em at the end of the plateau, but dips to below both SNe~1999em and 1999gi during the tail phase. Although the decline from the plateau to the tail is greater, and steeper, in SN~2003gd than in the normal type II-P supernovae, the tail luminosity is much brighter than that of the faint, nickel-poor SN~1997D.

\begin{table}
\begin{center}
\caption[]{Distance estimates of NGC 3184 found in literature.}
\begin{tabular}{lrrr}\hline
\label{tab:99gi}
Method & Source & Distance (Mpc) & Mean\\
\hline
Tully-Fisher & 1 & 14.9 & 11.1\\
             & 2 & 7.2  &     \\
Tertiary indicators & 3 & 7.9 & 7.9\\
EPM & 4 & 11.1 & 11.1\\
Kinematic & 1 & 10.6 & 9.9\\
          & 5 & 9.1 &\\
\hline
Mean          &  &  & $10.0\pm0.8$\\
\hline
\end{tabular}
\end{center}
1~http://leda.univ-lyon1.fr, 2~\citet{1994ApJ...430...53P}, 3~\citet{1979ApJ...227..729D}, 4~\citet{2002AJ....124.2490L}, 5~\citet{1988ngc..book.....T}\\
\end{table}

\begin{figure}
\begin{center}
\epsfig{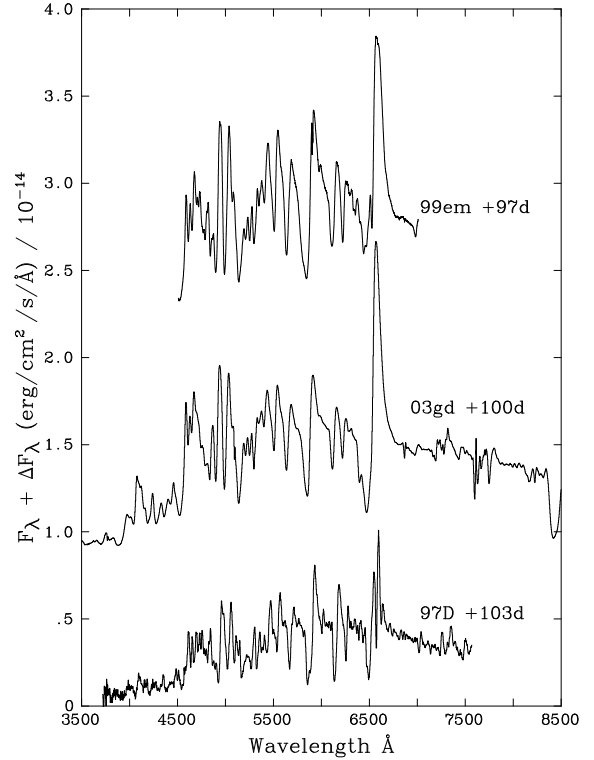}
\caption{Comparison between the spectra of SNe~2003gd, 1999em and 1997D which have nickel masses of 0.016, 0.048 \citep{2003ApJ...582..905H} and 0.006 \citep{2004MNRAS.347...74P} \msun, respectively, corrected for the distances used in this paper. SNe~1999em and 1997D are representative of a ``normal'' and a ``faint'' type II-P supernova respectively.}
\label{fig:99em03gd97D}
\end{center}
\end{figure}

The spectrum of SN~2003gd at 100 days is compared to those of SNe~1999em and 1997D at a similar phase, in Figure \ref{fig:99em03gd97D}. The spectra have been flux scaled using the method of \S\ref{sec:spec}, with the spectra of 1997D being scaled by an extra factor of five for clarity. The spectrum of SN~2003gd shows some similarities to both SNe~1997D and 1999em, but more so to that of SN~1999em with its broad spectral features. The spectral flux of SN~1997D is much fainter than that of SN~2003gd, and also has much narrower spectral features indicative of the lower velocities. The appearance of the spectra, the higher velocities, the colour evolution and higher plateau luminosities of SN~2003gd suggest that it is more akin to the ``normal'' type II-P events.

\subsection{Implications for the progenitor of SN~2003gd}

\citet{2004Sci...303..499S} present the discovery of a red supergiant which exploded as SN~2003gd. The galaxy was imaged six to nine months before the supernova explosion by {\it HST} and the Gemini Telescope. {\it HST} images after the explosion confirmed the positional coincidence of the supernova with a single resolved star. The observational properties of the star were compared with stellar evolutionary tracks. The progenitor star's colours and luminosity were found to be consistent with those of a red supergiant with an initial main sequence mass of $8^{+4}_{-2}$ \msun. However, the reddening and distance used in \citet{2004Sci...303..499S} has since been updated to $A_V = 0.43\pm0.19$ (\S\ref{sec:red}) and $D = 9.3\pm1.8$ (\S\ref{sec:D}), and we recalculate the bolometric luminosity of the progenitor star using these values. We find $M_{\rm bol} = -6.0\pm 0.7$ with no difference in the corresponding luminosity of $\log (L/{\rm L}_{\odot}) = 4.3\pm 0.3$. We therefore find no changes to the authors' conclusions.

We use Equation 2 of \citet{2003MNRAS.346...97N}, shown here in Equation \ref{equ:M},
\begin{eqnarray}
\label{equ:M}
\log \left(\frac{M_{\rm ej}}{{\rm M}_\odot}\right) &=& 0.234 M_{\rm V}+2.91\log \left(\frac{\Delta t}{\rm days}\right)\nonumber\\
& + &1.96 \log \left(\frac{u_{\rm ph}}{10^3\; {\rm km\:s}^{-1}}\right)-1.829
\end{eqnarray}
to compare the observed properties of SN~2003gd with the observed mass of $8^{+4}_{-2}$ \msun. Since we can only make assumptions about the length of the plateau, we have used this equation to constrain the plateau length to ascertain whether it is consistent with other type II-P SNe. Based on the observed mass we have estimated the ejected mass to be 6 \msun, leaving 2 \msun\ for the mass of the compact remnant plus other mass losses from the system. We find that the observed mass is consistent with a plateau length of $67^{+34}_{-25}$ days, where the errors are the combined errors in each of the parameters in the equation. For the error in the ejected mass we have added, in quadrature, an extra error of 1 \msun\ to the error in the observed mass. Using the full range of the observed mass, and other parameters, we find the supernova parameters are consistent with a plateau range of 40-119 days, although realistically this is unlikely to be below around 90 days, from what is commonly observed for SNe~II-P. Within the errors this mass appears not to be inconsistent with the observations and may suggest a mass in the upper end of the mass range.

The model of \citet{2003MNRAS.338..711Z}, when applied to SN~2003gd (see \S\ref{sec:zamp}), also seems to favour a main sequence mass in the upper end of the observed range, although this is achieved with a distance in the lower end of our distance range (Model A). Model B estimates an ejected mass of 13~\msun\ using 9.2 Mpc, so is directly comparable with the observed mass. This is just outside the observed mass range, but is not necessarily inconsistent given the theoretical uncertainties associated with the model. The model does appear to rule out an explosion date closer to discovery (Model C). We would like to stress though that more observations of progenitors are needed to fully test and constrain this model.

The observations of progenitors have important implications for the low-luminosity ``faint'' SNe~II-P. There are two very different plausible progenitor models for these, one being the low-energy explosion of massive stars \citep[e.g.][]{1998ApJ...498L.129T,2001MNRAS.322..361B,2003MNRAS.338..711Z}. In this model the collapsing core forms a black hole and a significant amount of fall-back of material occurs. An alternative scenario is the low-energy explosion of low-mass stars, presented by \citet{2000A&A...354..557C} who successfully reproduced the observations of SN~1997D with an explosion of $10^{50}$ ergs and an ejected mass of 6 \msun. The findings of \citet{astro-ph/0310057} support the high-mass progenitor scenario. The authors find a bimodal distribution of supernovae in the nickel mass, initial main sequence mass ($M_{\rm MS}$) plane. A reproduction of the authors' Figure 1 (right) is shown here in Figure \ref{fig:zamp} with the observational points of SNe~2003gd and 1999em added with filled circles. Zampieri remarks that the theoretical estimates of $M_{\rm MS}$ from \citet{astro-ph/0310057}, shown with open circles in the figure, will most probably need to be revised using the value of the opacity from \S\ref{sec:zamp}. In this picture the upper branch forms the ``normal'' type II-P branch and the lower the ``faint'' branch, with SN~2003gd occupying the lower end of the ``normal'' branch. The alternative low-mass scenario positions the ``faint'' SNe at the lower end of the ``normal'' branch in one continuous distribution. Until there are direct detections of the progenitor stars of these supernovae, we cannot determine if either of these models is a true model.

SNe~II-P produce a large range of nickel masses from around $10^{-3}$ to $10^{-1}$ \msun. \citet{2003ApJ...582..905H} showed that a correlation exists between the amount of nickel produced in the explosion and the absolute magnitude in the $V$-band during the plateau phase. A reproduction of \citeauthor{2003ApJ...582..905H}'s Figure 3 is shown here in Figure \ref{fig:hamuy} with SN~2003gd shown with a filled triangle. Four other supernovae, SNe~1994N, 1999eu, 2001dc and 2003Z represented with open stars, are from \citet{2003PhDT}. These four supernovae plus SN~1999br are from the group of ``faint'' type II-P supernovae. In this figure, SN~2003gd is in the centre of a continuous distribution of supernovae between the low nickel, faint SN~1999br and the high nickel, bright SN~1992am.

\citet{2003MNRAS.343..735S} made a comparison between all the known, direct information available on supernova progenitors with the theoretical predictions of \citet{2003ApJ...591..288H}, whose work on the pre-supernova evolution of massive stars produced supernova population diagrams as a function of mass and metallicity. These models do not consider binary stars or rotation. We plot this supernova population diagram in Figure \ref{fig:heger} and add SN~2003gd, which has a solar metallicity \citep{2004Sci...303..499S}. We repeat the caution which \citeauthor{2003MNRAS.343..735S} issues, emphasising that at present this is a qualitative comparison. Whilst saying this, the progenitor detected for SN~2003gd is consistent with the theoretical II-P region undergoing a O/Ne/Mg core collapse albeit close to the theoretical boundaries.

\begin{figure}
\begin{center}
\epsfig{file = Zampieri.epsi, angle = -90, width = 83mm}
\caption{Reproduction of Figure 1 (right) of \citet{astro-ph/0310057} showing the bimodal distribution of type II-P supernovae, in the $M_{\rm Ni}/M_{\rm MS}$ plane, from their model. The open circles show theoretical data from the semi-empirical model of \citet{2003MNRAS.338..711Z}, where $M_{\rm MS}$ is estimated from the ejected envelope mass. The filled circles show observational data for SNe~2003gd and 1999em, where $M_{\rm Ni}$ is estimated from the light curve  and $M_{\rm MS}^{99em}$ is an upper mass limit only \citep{2002ApJ...565.1089S}. SN~2003gd inhabits the lower end of the ``normal'' type II-P branch.}
\label{fig:zamp}
\end{center}
\end{figure}

\begin{figure}
\begin{center}
\epsfig{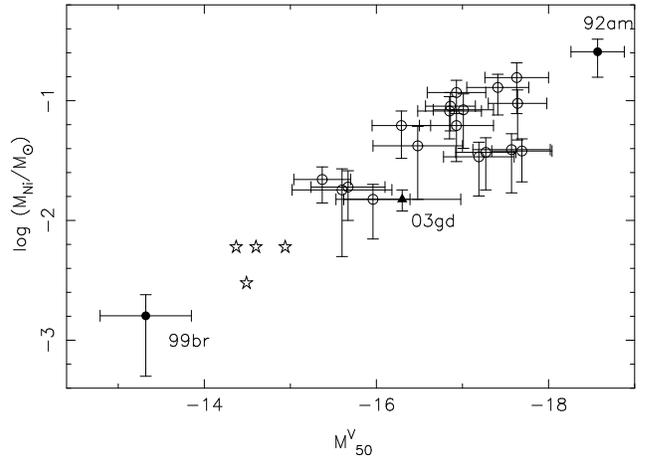}
\caption{Reproduction of Figure 3 of \citet{2003ApJ...582..905H} showing the correlation between the mass of $^{56}$Ni synthesised in the supernova with the absolute magnitude in the $V$-band at 50 days after explosion. The figure has additional points of SN~2003gd, shown with a filled triangle, and four ``faint'' SNe~1994N, 1999eu, 2001dc and 2003Z, shown with open stars, from \citet{2003PhDT}.}
\label{fig:hamuy}
\end{center}
\end{figure}

\begin{figure*}
\begin{center}
\epsfig{file = 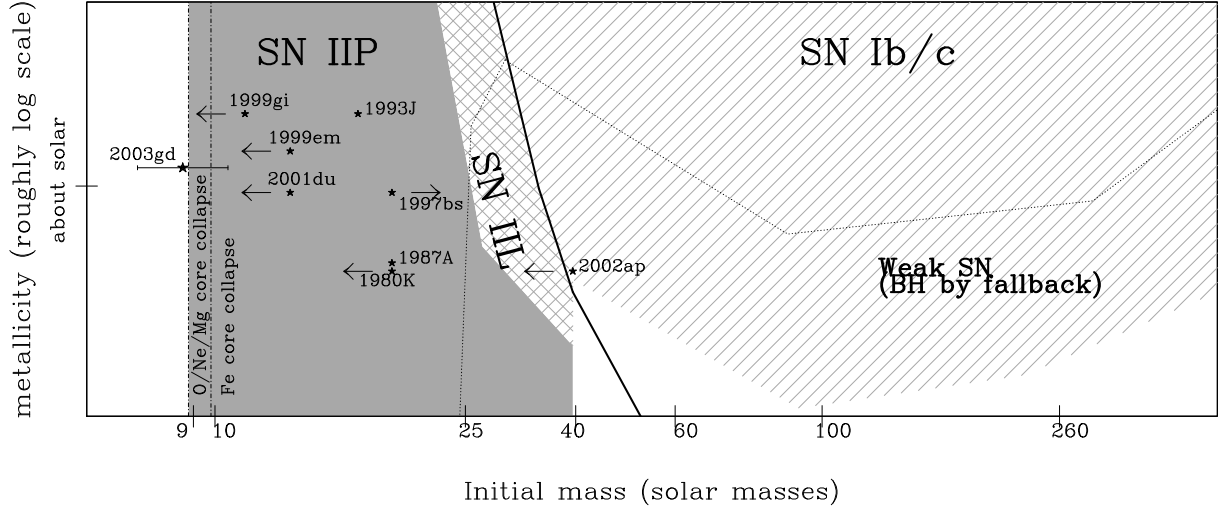, angle = -90, width = 170mm}
\caption{Theoretical SNe populations from the models of \citet{2003ApJ...591..288H}. All objects below the dotted line should produce low luminosity SNe due to black hole formation and subsequent fall-back of material onto the iron core. All objects to the right of the solid black line should have lost their hydrogen envelope by the point of collapse resulting in hydrogen-free SNe. The white regions indicate areas where SN should not be observed. The observed positions of the SNe are approximately estimated and placed accordingly, where the arrows indicate those with limits. The zero point on the y-axis is assumed to be $\log (Z/{\rm Z}_{\odot}) = -1$.}
\label{fig:heger}
\end{center}
\end{figure*}

\section{conclusions}\label{sec:con}

We have presented photometric and spectroscopic data on the type II-P SN~2003gd. We compared the $BVRI$ light curves with those of the well observed, spectroscopically similar SN~1999em using a $\chi^2$-fitting algorithm. This analysis allowed us to estimate an explosion epoch of JD 2452717$\pm$21, corresponding to a date of 18th March 2003, and the $VI$ magnitudes during the plateau phase. It was noted during this analysis that the tail luminosity of SN~2003gd was much lower than that of SN~1999em. Using this explosion epoch we were able to compare the velocity evolution of SN~2003gd with those of other similar type II-P SNe. The velocity evolution was found to be comparable at all epochs in our dataset. This enabled us to extrapolate the velocity evolution of SN~2003gd backwards to find the velocity during the plateau phase.

We calculated three distances to M74 using two different methods, the Standardised Candle Method \citep[SCM;][]{2003ApJ...582..905H,2004mmu..sympE...2H,astro-ph/0309122} and the Brightest Supergiant Method \citep[BSM;][]{1994MNRAS.271..530R,1994A&A...286..718K}. Firstly, using the inferred velocity and $VI$ magnitudes, we estimated a distance of $D = 9.6\pm 1.8$ Mpc with the SCM. We then used {\it HST} (WFPC2) photometry to estimate the distances of $7.7\pm 1.7$ Mpc and $9.6\pm 2.2$ Mpc using two different BSM calibrations. Using these three distances and other distances within the literature we estimated an overall distance of $9.3\pm 1.8$ Mpc to M74. This distance allowed the mass of $^{56}$Ni to be estimated at 0.016~\msun.

A comparison was made between the observed data of SN~2003gd and the semi-analytical models of \citet{2003MNRAS.338..711Z}. These models firstly tested the input parameters of the code and secondly ascertained whether the distance and explosion epoch were consistent with the observed progenitor mass. It was found that in order for the model to produce an ejected mass consistent with the observed progenitor mass the opacities needed were large, $\kappa = 0.3-04$ cm$^{-2}$ g$^{-1}$, suggesting little mixing occurred in SN~2003gd. The results of the models suggested that the progenitor was probably in the upper end of the progenitor mass range of $8^{+4}_{-2}$ \msun\ and the explosion was not likely to have occurred much later than estimated in \S\ref{sec:epoch}. 

The low tail luminosity of SN~2003gd sparked a debate about the nature of this supernova, whether it was a ``normal'' type II-P or whether it belonged to the group of ``faint'', low nickel SNe. To ascertain this we compared the colour evolution, pseudo-bolometric light curve and a spectrum (100 days) of SN~2003gd with those of SNe belonging to the ``normal'' and ``faint'' groups. We also compared the observed mass of the supernova with theoretical models. Using all the information available for SN~2003gd we conclude the following:
\begin{enumerate}
\item The colour evolution of SN~2003gd is similar to that of the ``normal'' type II-P SNe, although the reddening at the end of the plateau phase is slightly steeper it does not show a colour excess like the ``faint'' supernovae.
\item The plateau luminosity of SN~2003gd is comparable to those of normal type II-P SNe although the decline from the plateau to the tail is greater.
\item The tail luminosity of SN~2003gd, despite the larger decrease, is greater than the tail luminosity of the ``faint'' type II-P SNe.
\item The spectra are much more similar to the ``normal'' type II-P SNe, with their broad spectral features, than the ``faint'' SNe, which are much fainter and have much narrower spectral features.
\item The velocities of SN~2003gd are comparable to those of the normal type II-P group.
\item SN~2003gd lies in the centre of the correlated, continuous distribution of SNe~II-P in the $M_{\rm Ni}$/$M_{\rm V}$ plane which has the low nickel, faint SN~1999br and high nickel, bright SN~1992am at its extremities.
\item The observed progenitor mass of SN~2003gd lies at the lower end of the ``normal'' type II-P SN branch in the $M_{\rm Ni}$/$M_{\rm MS}$ plane \citep{2003MNRAS.338..711Z}.
\item In the SN population diagram of \citet{2003ApJ...591..288H} the observed progenitor mass of SN~2003gd lies in the type II-P region, which undergoes O/Ne/Mg core collapse, albeit close to the boundaries.
\end{enumerate}
We therefore conclude that the observations and observed properties of SN~2003gd are consistent with the observed progenitor star, a red supergiant of mass $8^{+4}_{-2}$ \msun\ \citep{2004Sci...303..499S}, which resulted in a normal type II-P supernova at the end of its life.
\\
\\
\section*{Acknowledgements}
Observations were made using the WHT, INT and JKT, which are operated by the Isaac Newton Group (ING), and also with the Italian TNG, which is operated by the Centro Galileo Galilei of the INAF (Instituto Nazionale di Astrofisica). These telescopes are operated on the island of La Palma at the Spanish Observatorio del Roque de los Muchachos of the Instituto de Astrof\'{\i}sica de Canarias. This paper is also partially based on observations collected at the Asiago Observatory, Italy, and at the European Southern Observatory, Chile. We would like to thank the ING for their rapid response to our service proposal, and D. Lennon and F. Prada for making these observations. We would also like to thank Simona Di Tomaso, Gloria Andreuzzi and the staff at the TNG who contributed to taking the TNG observations and for their excellent support. We would also like to thank Fernando Patat for obtaining the VLT data. M. Hendry, J. Maund and S. Smartt would like to thank PPARC for their financial support. S. Leon and S. Verley are partially supported by the Spanish MCyT Grant AYA 2002-03338 and S. Leon by an Averroes Fellowship from the Junta de Andaluc\'{\i}a.

\label{lastpage}
\end{document}